\documentclass[a4paper,11pt]{article}
\pdfoutput=1 

\usepackage{jheppub} 
\makeatletter
\DeclareRobustCommand*{\bfseries}{%
  \not@math@alphabet\bfseries\mathbf
  \fontseries\bfdefault\selectfont
  \boldmath
}
\makeatother

\usepackage{calc}
\usepackage{rotating}
\usepackage[english]{babel}
\usepackage[utf8]{inputenc}
\usepackage{graphicx}
\usepackage{subfig}
\usepackage{float}
\usepackage{amsmath}
\usepackage{amssymb}
\usepackage{amsthm}
\usepackage{latexsym}
\usepackage{dcolumn}
\usepackage{hyperref}
\newcommand{\newc}{\newcommand*}

\long\def\begincomment#1\endcomment{%
        \begingroup\sf\baselineskip12pt#1\endgroup}

\newc{\etal}{\textrm{et al.}} 
\newc{\eg}{\textrm{e.g.}} 
\newc{\ie}{\textrm{i.e.}}
\newc{\etc}{\textrm{etc}}
\newc\vs{\textrm{vs.}}
\newc{\cl}{\rm {C.L.}}

\newc{\ev}{\ensuremath{\,\mathrm{eV}}}
\newc{\kev}{\ensuremath{\,\mathrm{keV}}}
\newc{\mev}{\ensuremath{\,\mathrm{MeV}}}
\newc{\gev}{\ensuremath{\,\mathrm{GeV}}}
\newc{\tev}{\ensuremath{\,\mathrm{TeV}}}
\newc{\MeV}{\mev} 
\newc{\TeV}{\tev}
\newc{\invpb}{\ensuremath{/\text{pb}}}
\newc{\invfb}{\ensuremath{/\text{fb}}}
\newc\nb{\ensuremath{\,\mathrm{nb}}} \newc\pb{\ensuremath{\,\mathrm{pb}}} \newc\fb{\ensuremath{\,\mathrm{fb}}}
\newc\pc{\ensuremath{\,\mathrm{pc}}}
\newc\kpc{\ensuremath{\,\mathrm{kpc}}}
\newc\mpc{\ensuremath{\,\mathrm{Mpc}}}
\newc\ps{\ensuremath{\,\mathrm{ps}}} 
\newc\cmeter{\ensuremath{\,\mathrm{cm}}} 
\newc\meter{\ensuremath{\,\mathrm{m}}} 
\newc\kmeter{\ensuremath{\,\mathrm{km}}}
\newc\second{\ensuremath{\,\mathrm{s}}}
\newc\msecond{\ensuremath{\,\mathrm{ms}}}
\newc\nsecond{\ensuremath{\,\mathrm{ns}}}
\newc\psecond{\ensuremath{\,\mathrm{ps}}}

\newc{\chisqmin}{\ensuremath{\chi^2_{\mathrm{min}}}}
\newc{\Delchisq}{\ensuremath{\Delta\chi^2}}
\newc{\chisq}{\ensuremath{\chi^2}}
\newc{\like}{\ensuremath{\mathcal{L}}}

\newc\lsim{\ensuremath{\mathrel{\rlap{\lower4pt\hbox{\hskip1pt$\sim$}}\raise1pt\hbox{$<$}}}}
\newc\gsim{\ensuremath{\mathrel{\rlap{\lower4pt\hbox{\hskip1pt$\sim$}}\raise1pt\hbox{$>$}}}}
\newc{\VEV}[1]{\ensuremath{\langle #1 \rangle}}
\newc{\dl}{\ensuremath{\stackrel{\leftarrow}{D}}}
\newc{\dr}{\ensuremath{\stackrel{\rightarrow}{D}}}

\newc{\bcenter}{\begin{center}}    \newc{\ecenter}{\end{center}}
\newc{\bfl}{\begin{flushleft}}    \newc{\efl}{\end{flushleft}}
\newc{\bfr}{\begin{flushright}}    \newc{\efr}{\end{flushright}}

\newc{\bi}{\begin{itemize}}
\newc{\ei}{\end{itemize}}
\newc{\bed}{\begin{description}}
\newc{\eed}{\end{description}}
\newc{\ben}{\begin{enumerate}}
\newc{\een}{\end{enumerate}}

\newc{\be}{\begin{equation}}
\newc{\ee}{\end{equation}}
\newc{\bea}{\begin{eqnarray}}
\newc{\eea}{\end{eqnarray}}
\newc{\bfle}{\begin{flalign}}
\newc{\efle}{\end{flalign}}
\newc{\ra}{\rightarrow}

\newc{\alphas}{\ensuremath{\alpha_s}}
\newc{\alphatwo}{\ensuremath{\alpha_2}}
\newc{\alphaone}{\ensuremath{\alpha_1}}
\newc{\alphai}[1]{\ensuremath{\alpha_{#1}}}
\newc{\alphaem}{\ensuremath{\alpha_{\mathrm{em}}}}
\newc{\alphaeff}{\ensuremath{\alpha_{\mathrm{eff}}}}
\newc{\sineff}{\ensuremath{\sin \theta_{\mathrm{eff}}}}
\newc{\sinsqeff}{\ensuremath{\sin^2 \theta_{\mathrm{eff}}}}
\newc{\dalphahad}{\ensuremath{\Delta \alpha_{\mathrm{had}}}}
\newc{\yt}{\ensuremath{h_t}} \newc{\yb}{\ensuremath{h_b}} \newc{\ytau}{\ensuremath{h_{\tau}}}
\newc\mz{\ensuremath{M_Z}} 
\newc\mw{\ensuremath{m_W}}
\newc\mZ{\mz}        \newc\mW{\mw}
\newc\mhsm{\ensuremath{ m_{H_{\mathrm{SM}}}}}
\newc{\mtop}{\ensuremath{ m_t}}               \newc{\mtpole}{\ensuremath{ M_t}}
\newc{\mbottom}{\ensuremath{ m_b}} 
\newc{\mtau}{\ensuremath{ m_{\tau}}}
\newc{\mt}{\mtpole}
\newc{\mb}{\mbottom} 
\newc{\rtwogg}{\ensuremath{R_{h_2}(\gamma\gamma)}}
\newc{\rtwozz}{\ensuremath{R_{h_2}(ZZ)}}
\newc{\ronegg}{\ensuremath{R_{h_1}(\gamma\gamma)}}
\newc{\ronezz}{\ensuremath{R_{h_1}(ZZ)}}
\newc{\rsiggg}{\ensuremath{R_{h_\textrm{sig}}(\gamma\gamma)}}
\newc{\rsigzz}{\ensuremath{R_{h_\textrm{sig}}(ZZ)}}
\newc{\llbar}{\ensuremath{\ell\bar{\ell}}}
\newc{\tauptaum}{\ensuremath{ \tau^+\tau^-}}
\newc{\qqbar}{\ensuremath{ q\bar{q}}} \newc{\ppbar}{\ensuremath{ p\bar{p}}}
\newc{\bbbar}{\ensuremath{ b\bar{b}}} \newc{\ttbar}{\ensuremath{ t\bar{t}}}
\newc{\ffbar}{\ensuremath{ f\bar{f}}} \newc{\tautaubar}{\ensuremath{ \tau\bar{\tau}}}

\newc{\mchi}{\ensuremath{m_\neutone}}
\newc{\squark}{\ensuremath{\tilde{q}}}
\newc{\slepton}{\ensuremath{\tilde{l}}}
\newc{\gluino}{\ensuremath{\tilde{g}}} 
\newc{\mgluino}{\ensuremath{{m_{\gluino}}}}
\newc{\wino}{\ensuremath{\tilde{W}}} 
\newc{\mwino}{\ensuremath{{m_{\wino}}}}
\newc{\tone}{\ensuremath{{\tilde{t}_1}}}
\newc{\Hone}{\ensuremath{{\tilde{H}_{1}}}}
\newc{\Htwo}{\ensuremath{{\tilde{H}_{2}}}}
\newc{\Hhtwo}{\ensuremath{{H_{2}}}}
\newc{\qli}{\ensuremath{{\tilde{Q}_{i}}}}
\newc{\uri}{\ensuremath{{\tilde{u}_{i}}}}
\newc{\dri}{\ensuremath{{\tilde{d}_{i}}}}
\newc{\lli}{\ensuremath{{\tilde{L}_{i}}}}
\newc{\eri}{\ensuremath{{\tilde{e}_{i}}}}

\newc{\sthw}{\ensuremath{ \sin\theta_W}}              \newc{\cthw}{\ensuremath{\cos\theta_W}}
\newc{\tanthw}{\ensuremath{ \tan\theta_W}}              \newc{\cotthw}{\ensuremath{\cot\theta_W}}
\newc{\ssqthw}{\ensuremath{\sin^2 \theta_W}}
\newc{\msbar}{\ensuremath{\overline{MS}}} \newc{\drbar}{\ensuremath{\overline{DR}}}
\newc{\mtmtsmmsbar}{\ensuremath{ m_t(m_t)^{\msbar}_{{\mathrm{SM}}}}}
\newc{\mtmtsmdrbar}{\ensuremath{ m_t(m_t)^{\drbar}_{{\mathrm{SM}}}}}
\newc{\mtmtmssmdrbar}{\ensuremath{ m_t(m_t)^{\drbar}_{{\mathrm{SUSY}}}}}
\newc{\mbmbmsbar}{\ensuremath{ m_b(m_b)^{\msbar} }}
\newc{\mbmbsmmsbar}{\ensuremath{ m_b(m_b)^{\msbar}_{{\mathrm{SM}}}}}
\newc{\mbmzsmmsbar}{\ensuremath{ m_b(\mz)^{\msbar}_{{\mathrm{SM}}}}}
\newc{\mbmzsmdrbar}{\ensuremath{ m_b(\mz)^{\drbar}_{{\mathrm{SM}}}}}
\newc{\mbmzmssmdrbar}{\ensuremath{ m_b(\mz)^{\drbar}_{{\mathrm{SUSY}}}}}
\newc{\mtaumzsmmsbar}{\ensuremath{ m_{\tau}(\mz)^{\msbar}_{{\mathrm{SM}}}}}
\newc{\mtaumzsmdrbar}{\ensuremath{ m_{\tau}(\mz)^{\drbar}_{{\mathrm{SM}}}}}
\newc{\mtaumzmssmdrbar}{\ensuremath{ m_{\tau}(\mz)^{\drbar}_{{\mathrm{SUSY}}}}}
\newc{\alphasmzms}{\ensuremath{\alpha_s(M_Z)^{\overline{MS}}}}
\newc{\alphaimzms}[1]{\ensuremath{\alpha_{#1}(M_Z)^{\overline{MS}}}}
\newc{\alphaemmz}{\ensuremath{\alpha_{\mathrm{em}}(M_Z)^{\overline{MS}}}}

\newc{\mzero}{\ensuremath{{m_0}}}
\newc{\mhalf}{\ensuremath{ m_{1/2}}}
\newc{\tanb}{\ensuremath{\tan\beta}}
\newc{\azero}{\ensuremath{ A_0}}
\newc{\signmu}{\ensuremath{\rm{sgn}\,\mu}}
\newc{\atau}{\ensuremath{{A_{\tau}}}}
\newc{\mueff}{\ensuremath{\mu_{\rm{eff}}}}
\newc{\lam}{\ensuremath{{\lambda}}}
\newc{\kap}{\ensuremath{{\kappa}}}
\newc{\alam}{\ensuremath{{A_{\lambda}}}}
\newc{\akap}{\ensuremath{{A_{\kappa}}}}
\newc{\hs}{\ensuremath{ H_s}}      
\newc{\mhs}{\ensuremath{ m_{H_s}}} 
\newc{\mgut}{\ensuremath{ M_{\rm GUT}}}
\newc{\mvl}{\ensuremath{ M_{\rm VL}}}
\newc{\gut}{\ensuremath{{\rm GUT}}}
\newc{\mplanck}{\ensuremath{ M_{\rm P}}}      \newc{\mpl}{\ensuremath{ M_{\rm Pl}}}
\newc{\msusy}{\ensuremath{ M_{\rm SUSY}}}      \newc{\ms}{\ensuremath{ M_{\rm S}}}
 \newc{\hu}{\ensuremath{ H_u}}       \newc{\hd}{\ensuremath{ H_d}}
 \newc{\mhu}{\ensuremath{ m_{H_u}}}       \newc{\mhd}{\ensuremath{ m_{H_d}}}
 \newc{\mhuew}{\ensuremath{ m^{\ast}_{H_u}}}       \newc{\mhdew}{\ensuremath{ m^{\ast}_{H_d}}}
 \newc{\mhuewsq}{\ensuremath{ m^{\ast\, 2}_{H_u}}}       \newc{\mhdewsq}{\ensuremath{ m^{\ast\, 2}_{H_d}}}
 \newc{\mhl}{\ensuremath{m_\hl}} 
 \newc{\mhone}{\ensuremath{m_{h_1}}} 
 \newc{\mhtwo}{\ensuremath{m_{h_2}}} 
 \newc{\mhi}{\ensuremath{m_{\tilde{h}}}} 
 \newc{\mul}{\ensuremath{m_{\tilde{u}_L}}} 
 \newc{\mtone}{\ensuremath{m_{\tilde{t}_1}}} 
 \newc{\ma}{\ensuremath{m_A}} 
 \newc{\mH}{\ensuremath{m_H}} 
 \newc{\maone}{\ensuremath{m_{a_1}}} 
 \newc{\matwo}{\ensuremath{m_{a_2}}}
 \newc{\hone}{\ensuremath{h_1}}
 \newc{\htwo}{\ensuremath{h_2}}
 \newc{\aone}{\ensuremath{a_1}}
 \newc{\atwo}{\ensuremath{a_2}}
 \newc{\mqthree}{\ensuremath{m_{\tilde{Q}_3}^2}}
 \newc{\muthree}{\ensuremath{m_{\tilde{u}_3}^2}}
 \newc{\mqli}{\ensuremath{m_{\tilde{Q}_{i}}}}
 \newc{\muri}{\ensuremath{m_{\tilde{u}_{i}}}}
 \newc{\mdri}{\ensuremath{m_{\tilde{d}_{i}}}}
 \newc{\mlli}{\ensuremath{m_{\tilde{L}_{i}}}}
 \newc{\meri}{\ensuremath{m_{\tilde{e}_{i}}}}
 \newc{\ts}{\ensuremath{T_{SUSY}}}

\newc{\sigsip}{\ensuremath{\sigma^{\rm SI}_{p}}}	\newc{\sigsin}{\ensuremath{\sigma^{\rm SI}_{n}}}
\newc{\sigsdp}{\ensuremath{\sigma^{\rm SD}_{p}}}	\newc{\sigsdn}{\ensuremath{\sigma^{\rm SD}_{n}}}
\newc{\sigsi}{\ensuremath{\sigma^{\rm SI}}}	\newc{\sigsd}{\ensuremath{\sigma^{\rm SD}}}
\newc{\abund}{\ensuremath{ \Omega h^2}}
\newc{\omegadm}{\ensuremath{ \Omega_{{\rm DM}}}}     \newc{\abunddm}{\ensuremath{ \Omega_{{\rm DM}} h^2}} 
\newc{\omegam}{\ensuremath{ \Omega_{{\rm m}}}}       \newc{\abundm}{\ensuremath{ \Omega_{{\rm m}} h^2}}
\newc{\omegab}{\ensuremath{ \Omega_{{\rm b}}}}	\newc{\abundb}{\ensuremath{ \Omega_{{\rm b}} h^2}}
\newc{\omegatot}{\ensuremath{ \Omega_{{\rm TOT}}}}
\newc{\omegacdm}{\ensuremath{ \Omega_{{\rm CDM}}}}   \newc{\abundcdm}{\ensuremath{ \Omega_{{\rm CDM}} h^2}}
\newc{\omegalambda}{\ensuremath{ \Omega_{\Lambda}}} \newc{\abundlambda}{\ensuremath{ \Omega_{\Lambda} h^2}}
\newc{\omegarad}{\ensuremath{ \Omega_{{\rm rad}}}}  \newc{\abundrad}{\ensuremath{ \Omega_{{\rm rad}} h^2}}
\newc{\rhocrit}{\ensuremath{ \rho_{\rm crit}}}
\newc{\rhochi}{\ensuremath{ \rho_{\chi}}}
\newc{\abunchi}{\ensuremath{\Omega_\chi h^2}}
\newc{\abundlsp}{\ensuremath{\Omega_{\rm LSP}h^2}}

\newc{\amu}{\ensuremath{ a_{\mu}}}        \newc{\amususy}{\ensuremath{ a_{\mu}^{\mathrm{SUSY}}}}
\newc{\amuexpt}{\ensuremath{ a_{\mu}^{\mathrm{expt}}}}        \newc{\amusm}{\ensuremath{ a_{\mu}^{\mathrm{SM}}}}
\newc\deltaamu{\ensuremath{\Delta a_{\mu}}} \newc{\deltaamususy}{\ensuremath{\delta a_{\mu}^{\mathrm{SUSY}}}}
\newc\gmtwo{\ensuremath{ (g-2)_{\mu}}} 
\newc{\deltagmtwomususy}{\ensuremath{\delta\left(g-2\right)_{\mu}^{\mathrm{SUSY}}}}
\newc{\deltagmtwomu}{\ensuremath{\delta\left(g-2\right)_{\mu}}}
\newc\BR{\ensuremath{\rm BR}}
\newc\bsgamma{\ensuremath{ b\rightarrow s \gamma }}
\newc\bxsgamma{\ensuremath{\overline{B}\rightarrow X_{s}\gamma}}
\newc\brbsgamma{\ensuremath{\BR\left(\bsgamma\right)}}
\newc\brbxsgamma{\ensuremath{\BR\left(\bxsgamma\right)}}
\newc\bsmumu{\ensuremath{B_s\to\mu^+\mu^-}}
\newc\brbsmumu{\ensuremath{\BR\left(B_s\to\mu^+\mu^-\right)}}
\newc\bdmmumu{\ensuremath{\overline{B}_d\to\mu^+\mu^-}}
\newc\bbbarmix{\ensuremath{\overline{B}_s\mbox{-}B_s}}      
\newc\delmbs{\ensuremath{\Delta M_{B_s}}}
\newc{\butaunu}{\ensuremath{B_u \rightarrow \tau \nu}}
\newc{\brbutaunu}{\ensuremath{\BR\left(B_u \rightarrow \tau \nu\right)}}

\newcommand*{\reftable}[1]{Table~\ref{#1}}         
       
\newcommand*{\reffig}[1]{Fig.~\ref{#1}}

        \newcommand*{\refeq}[1]{Eq.~(\ref{#1})}   

     \newcommand*{\refsec}[1]{Sec.~\ref{#1}}

\newcommand*{\neutone}{\ensuremath{\tilde{\chi}^0_1}}

\let\oldcite\cite
\renewcommand*{\cite}{~\oldcite}

\newcommand*{\hl}{\ensuremath{h}}

\restylefloat{figure}

\title{Naturally small Yukawa couplings from trans-Planckian asymptotic safety}

\author{Kamila Kowalska,}
\author{Soumita Pramanick}
\author{and Enrico Maria Sessolo}

\affiliation{National Centre for Nuclear Research,\\
Pasteura 7, 02-093 Warsaw, Poland}

\emailAdd{kamila.kowalska@ncbj.gov.pl}
\emailAdd{soumita.pramanick@ncbj.gov.pl}
\emailAdd{enrico.sessolo@ncbj.gov.pl}

\abstract{In gauge-Yukawa systems embedded in the framework of trans-Planckian asymptotic safety
we discuss the dynamical generation of arbitrarily small Yukawa couplings driven by the presence of a non-interactive infrared-attractive fixed point in the renormalization group flow.
Additional ultraviolet-attractive fixed points guarantee that the theory remains well defined up to an infinitely high scale. We apply this mechanism to the Yukawa couplings of the Standard Model extended with right-handed neutrinos, finding that asymptotically safe solutions in agreement with the current experimental determination of the masses and mixing angles exist for Dirac neutrinos with normal mass ordering. We generalize the discussion by applying the same mechanism to a new-physics model with 
sterile-neutrino dark matter, where we generate naturally the feeble Yukawa interaction required to reproduce via freeze-in the correct relic abundance.}

\begin{document}
\maketitle

\setcounter{footnote}{0}

\section{Introduction\label{sec:intro}}

In several instances of particle physics and cosmology, Nature seems to favor 
coupling sizes that are unnaturally small with respect to what would be expected purely on dimensional grounds. One very well known case is the gauge-hierarchy problem, which is
the fact that the sole dimensionful parameter of the Standard Model~(SM) Lagrangian, 
the Higgs scalar mass $\mu_H^2$, 
is some 34 orders of magnitude smaller than the Planck scale, $M_{\textrm{Pl}}^2$. Another example is the cosmological constant, which, if it is the driving force behind the present day acceleration of the Universe, appears to be 122 orders of magnitude smaller than an expected, natural value of the order of $M_{\textrm{Pl}}^2$.
Another still, perhaps less dramatically severe but nonetheless utterly puzzling,  
is the smallness of neutrino masses, which stems directly from the results of
neutrino-oscillation experiments and from the cosmological upper bound on the total active-neutrino mass. 
If the Higgs mechanism provides the origin of neutrino masses, as seems to be the case for all other fermions of the SM, their tiny value would imply that the neutrino Yukawa couplings are between seven and thirteen orders of magnitude smaller than the Yukawa couplings of the charged fermions of the SM.     

Interestingly, it was pointed out in recent years that some of the problems that seem to imply an unnaturally small
parameter may find a natural solution in asymptotically safe~(AS) quantum gravity\cite{inbookWS}. Following the 
development of functional renormalization group~(RG) techniques\cite{WETTERICH199390,Morris:1993qb} AS gravity has 
emerged in the last few decades as a powerful
framework for a Wilsonian description of the fundamental nature of quantum field theories. In several 
seminal studies\cite{Reuter:1996cp,Lauscher:2001ya,Reuter:2001ag,Manrique:2011jc} it was shown that
in the trans-Planckian regime the quantum fluctuations of the metric field may induce
an interactive fixed point 
of the RG flow of the Einstein-Hilbert effective action, which comprises the cosmological constant and the Ricci scalar.  
Results based on the minimal truncation were then extended to include gravitational effective operators of 
increasing mass dimension\cite{Lauscher:2002sq,Litim:2003vp,Codello:2006in,Machado:2007ea,Codello:2008vh,Benedetti:2009rx,Dietz:2012ic,Falls:2013bv,Falls:2014tra}, which seem to confirm the persistence of trans-Planckian fixed points, as does
introducing matter-field operators in the Lagrangian.  
An ambitious program has taken shape around the intriguing possibility that the full system of gravity and matter may be proven to be non-perturbatively renormalizable\cite{Robinson:2005fj,Pietrykowski:2006xy,Toms:2007sk,Tang:2008ah,Toms:2008dq,Rodigast:2009zj,Zanusso:2009bs,Daum:2009dn,Daum:2010bc,Folkerts:2011jz,Oda:2015sma,Eichhorn:2016esv,Christiansen:2017gtg,Hamada:2017rvn,Christiansen:2017cxa,Eichhorn:2017eht}. 

In the context of AS gravity, the gauge hierarchy problem may for example be solved through 
the ``resurgence mechanism''\cite{Wetterich:2016uxm}, see also Refs.\cite{Eichhorn:2017als,Pawlowski:2018ixd}.  
Using functional RG techniques it was shown that, for some choices of the action truncation,
quantum gravity can induce a large positive trans-Planckian anomalous dimension 
of the Higgs mass parameter. This modifies the scaling behavior of the adimensional parameter $\tilde{\mu}_H^2\equiv \mu^2_H/k^2$, which ends up featuring
an \textit{irrelevant} fixed point at $\tilde{\mu}_H^{2 \ast}=0$. If, at some large trans-Planckian scale, $\tilde{\mu}_H^{2}$ is (for some reason) of order one, it rapidly decreases in its RG flow to the infrared~(IR), thus acquiring at the Planck scale a small value without fine tuning (it ``resurges'' then again towards large values once gravity decouples).  

In similar fashion, a trans-Planckian 
irrelevant direction associated with a null fixed-point may be at the origin of the smallness of the cosmological constant, as was shown for example in Ref.\cite{Falls:2014zba} in a toy model where only the conformal mode of the metric was quantized. More in general, scalar field theories 
coupled to quantum gravity have been shown in many instances to develop 
an irrelevant direction associated with a small fixed-point value of the quartic coupling\cite{Shaposhnikov:2009pv,Eichhorn:2017als,Reichert:2019car,Kwapisz:2019wrl,Eichhorn:2019dhg,Eichhorn:2020kca,Eichhorn:2020sbo,Eichhorn:2021tsx}.
In this context, it was 
an early phenomenological success of AS gravity coupled to the SM to show
the emergence of such an IR-attractive fixed point in 
the beta function of the Higgs quartic coupling, which led to a fairly accurate 
prediction for the Higgs boson mass\cite{Shaposhnikov:2009pv} years before its discovery at the LHC.

From a phenomenological perspective, however, perhaps what makes the connection between AS gravity and the SM most exciting is
the early observation that an ultraviolet (UV) fixed point 
of gravitational origin can cure the pathological high-energy behavior of the hypercharge gauge coupling\cite{Harst:2011zx,Christiansen:2017gtg,Eichhorn:2017lry}, while the gauge couplings of color and isospin remain asymptotically free\cite{Daum:2009dn,Daum:2010bc,Folkerts:2011jz}. The fact that the SM 
coupled to gravity may feature interactive UV fixed points 
bears important consequences for its predictivity, as it implies that 
the actual number of free parameters in the theory can be effectively cut down. 
The aforementioned hypercharge gauge coupling, or the top Yukawa coupling\cite{Eichhorn:2017ylw} may turn out to 
be predictable when coupled to gravity, if, once again, they feature an irrelevant direction of the trans-Planckian flow near 
their fixed point. 

Unfortunately, an explicit calculation of the quantum gravity contribution to the matter beta functions of the full SM remains a formidable endeavor, marred by very large theoretical uncertainties. To bypass the theoretical hurdles without renouncing the predictive power of the fixed-point analysis, some recent studies\cite{Eichhorn:2018whv,Alkofer:2020vtb} have adopted 
a more effective approach, based on a parametric description of AS gravitational interactions with matter, which allows one to ``guess'' the strength of the gravitational coupling on the assumption that the fixed points of the matter sector
should not be in tension with the low-scale phenomenology.  
The same effective 
approach was also recently employed to boost the predictivity of certain models of New Physics (NP)
for which only incomplete information on their masses and couplings can be extracted experimentally\cite{Wang:2015sxe,Grabowski:2018fjj,Kwapisz:2019wrl,Reichert:2019car,Kowalska:2020gie,Domenech:2020yjf,Kowalska:2020zve}.

Incidentally, an intriguing byproduct of the effective approach to coupling the SM with AS gravity is that it has allowed for the extraction of some flavor-related predictions\cite{Eichhorn:2018whv,Alkofer:2020vtb,Kowalska:2020gie}, 
despite gravity being flavor-blind by construction. 
This is because, on the one hand, when trying to match the trans-Planckian flow to IR observations one has to engage with the
nontrivial interplay of relevant and irrelevant fixed-point directions
of the Yukawa and fermion mixing matrix elements, which can lead in some occasions to the appearance of specific hierarchies and/or textures (this was noted very early in the SM, \textit{e.g.}, in Refs.\cite{Pendleton:1980as,Hill:1980sq}). 
On the other hand, one finds that flavor hierarchies
may be preserved under the RG flow if the gauge symmetry and particle content prevent large 
additive terms to the beta functions of the Yukawa matrices (moreover, in the quark sector, preserving flavor hierarchies is facilitated by the near diagonality of the Cabibbo-Kobayashi-Maskawa~(CKM) matrix). 

In this work, we explore in some generality the issue of
the smallness of the neutrino Yukawa couplings in the context of asymptotic safety.
We do so by translating 
to the lepton sector of the SM augmented by right-handed neutrinos (and to the Pontecorvo-Maki-Nakagawa-Sakata~(PMNS) matrix) the effective approach introduced for the quarks in Ref.\cite{Alkofer:2020vtb}.  
Our main goal is to investigate to what extent arbitrarily small coupling sizes can be dynamically generated without introducing \textit{ad hoc} fine tuning, neither among the parameters of the SM Lagrangian, nor in the parametric description of the AS trans-Planckian interactions. As can be foreseen, this is achieved in a similar fashion to the ``resurgence'' mechanism, by letting the trans-Planckian flow of an irrelevant parameter get arbitrarily close to its null fixed point before decoupling the gravitational interaction at the Planck scale. However, unlike in 
the resurgence mechanism, the flow of the (eventually small) Yukawa coupling remains here finite up to an infinitely high scale, due to the additional presence of UV-attractive fixed points. The size of the small parameter at decoupling is determined by the value of an integration constant of the flow, which delineates the position of the Planck scale along the trajectory conjoining those UV fixed points to the null irrelevant point in the IR.

We reiterate that we do not attempt to address in this work the many unresolved theoretical issues still associated with the general framework of AS gravity. Much work in this direction is currently undertaken in the community -- see, \textit{e.g.}, Refs.\cite{Donoghue:2019clr,Bonanno:2020bil}. 
Our goal is rather that of studying the implications of the AS framework for low-energy Lagrangians characterized by tiny Yukawa couplings. 
To this regard, we show that our findings extend beyond the physics of SM neutrinos by additionally considering a model where a sterile neutrino plays the role of dark matter. The relic abundance is obtained 
via the freeze-in mechanism\cite{McDonald:2001vt,Choi:2005vq,Kusenko:2006rh,Petraki:2007gq,Hall:2009bx} in the early Universe, which requires feeble Yukawa-like interactions of the dark matter particle with a heavy scalar field.  

The paper is organized as follows. In \refsec{sec:neutrino} we briefly review the observational status of neutrino masses 
and mixings, which provide the phenomenological background for the subsequent discussion. Section~\ref{sec:fpan}, dedicated to the dynamical mechanism for generating a small Yukawa coupling without fine tuning from asymptotic safety, is divided in three parts. In \refsec{sec:as} we remind the reader of some broadly known notions in AS gravity; in \refsec{sec:gen_mech} we explain in some generality how to obtain dynamically a small Yukawa coupling at the Planck scale; in \refsec{sec:top-neu} we exemplify the main features of this mechanism by applying it to a simplified model composed of only one neutrino generation and the top quark and we quantify the fine tuning of the gravitational effective coupling in different regions of the parameter space. 
We perform a full trans-Planckian 
fixed-point analysis of the SM lepton sector plus three right-handed neutrinos in \refsec{sec:SM}. We extend the discussion to a model of dark matter with feeble Yukawa interactions with the visible sector in \refsec{sec:freezein}, and we finally summarize our findings and conclude in \refsec{sec:summary}. We dedicate the appendix to the explicit form of the one-loop RGEs used in this work.

\section{Current status of neutrino masses and mixing\label{sec:neutrino}}

The observation of neutrino oscillations in various experiments\cite{Super-Kamiokande:2001ljr,SNO:2002tuh,Super-Kamiokande:1998kpq,KamLAND:2002uet,KamLAND:2004mhv,K2K:2002icj,MINOS:2006foh} indicates that neutrinos have a mass, which is much smaller than the masses of the other fermions. The most recent results of the global fit for the neutrino mass differences squared from NuFIT5.1 (2021)\cite{Gonzalez-Garcia:2012hef, Esteban:2020cvm} read
\begin{eqnarray}\label{mass_fit}
\Delta m_{21}^2 &=& 7.42^{+0.21}_{-0.20}\times 10^{-5} \, {\rm eV}^2,\label{mass_fit_2} \\
\textrm{NO:}\quad \Delta m_{31}^2 &=& 2.515^{+0.028}_{-0.028}\times 10^{-3} \, {\rm eV}^2,\label{mass_fit_31} \\
\textrm{IO:}\quad \Delta m_{32}^2 &=& -2.498^{+0.028}_{-0.029}\times 10^{-3} \, {\rm eV}^2, \label{mass_fit_32}
\end{eqnarray}
where $\Delta m_{31}^2>0$ refers to normal ordering (NO), while $\Delta m_{32}^2<0$ to inverted ordering (IO). Additionally, the 95\%~C.L.  upper bound on the sum of neutrino masses from cosmological measurements by the Planck satellite is\cite{Planck:2018vyg}
\be\label{planck}
\sum_{i=1,2,3} m_i < 0.12\,\textrm{eV}.
\ee

\begin{figure}[t]
\centering
\subfloat[]{%
\includegraphics[width=0.47\textwidth]{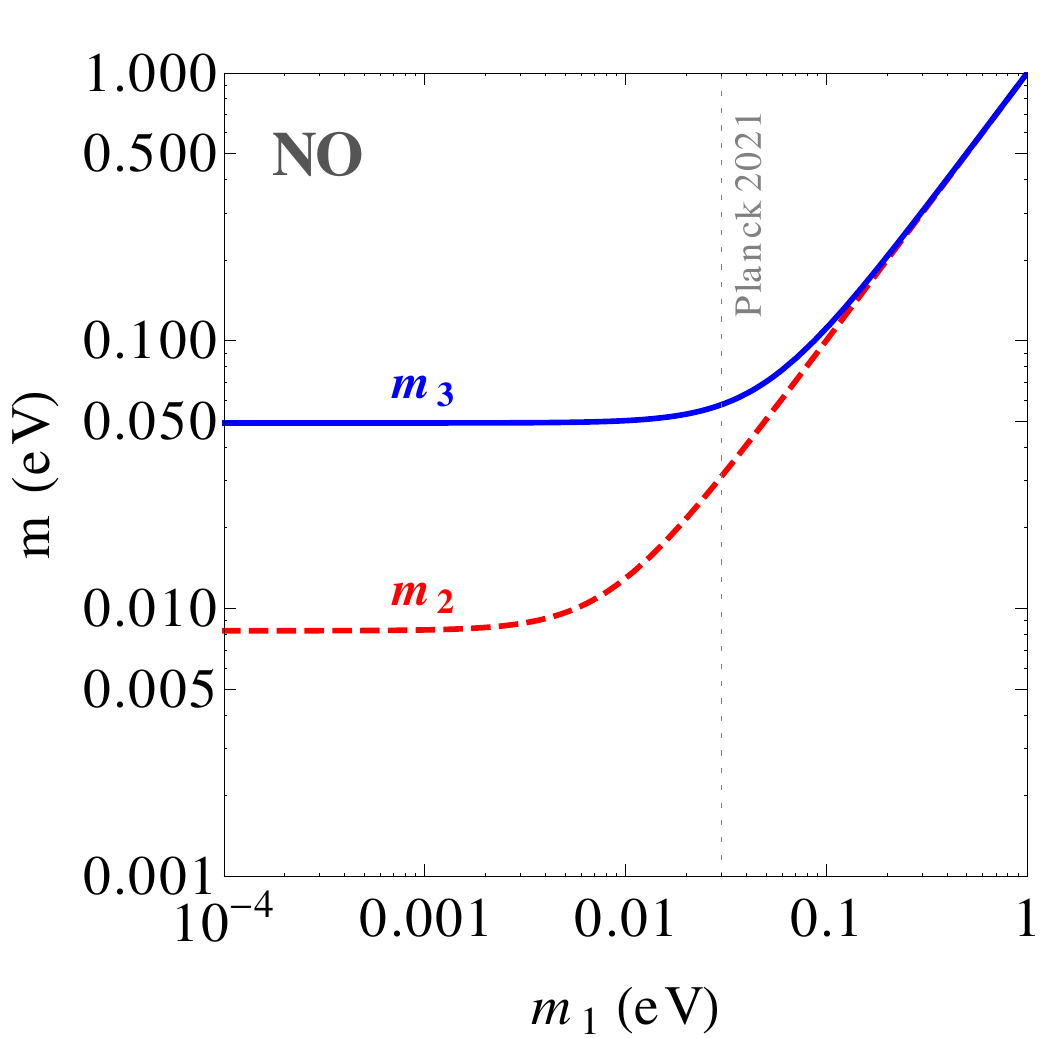}
}%
\hspace{0.02\textwidth}
\subfloat[]{%
\includegraphics[width=0.47\textwidth]{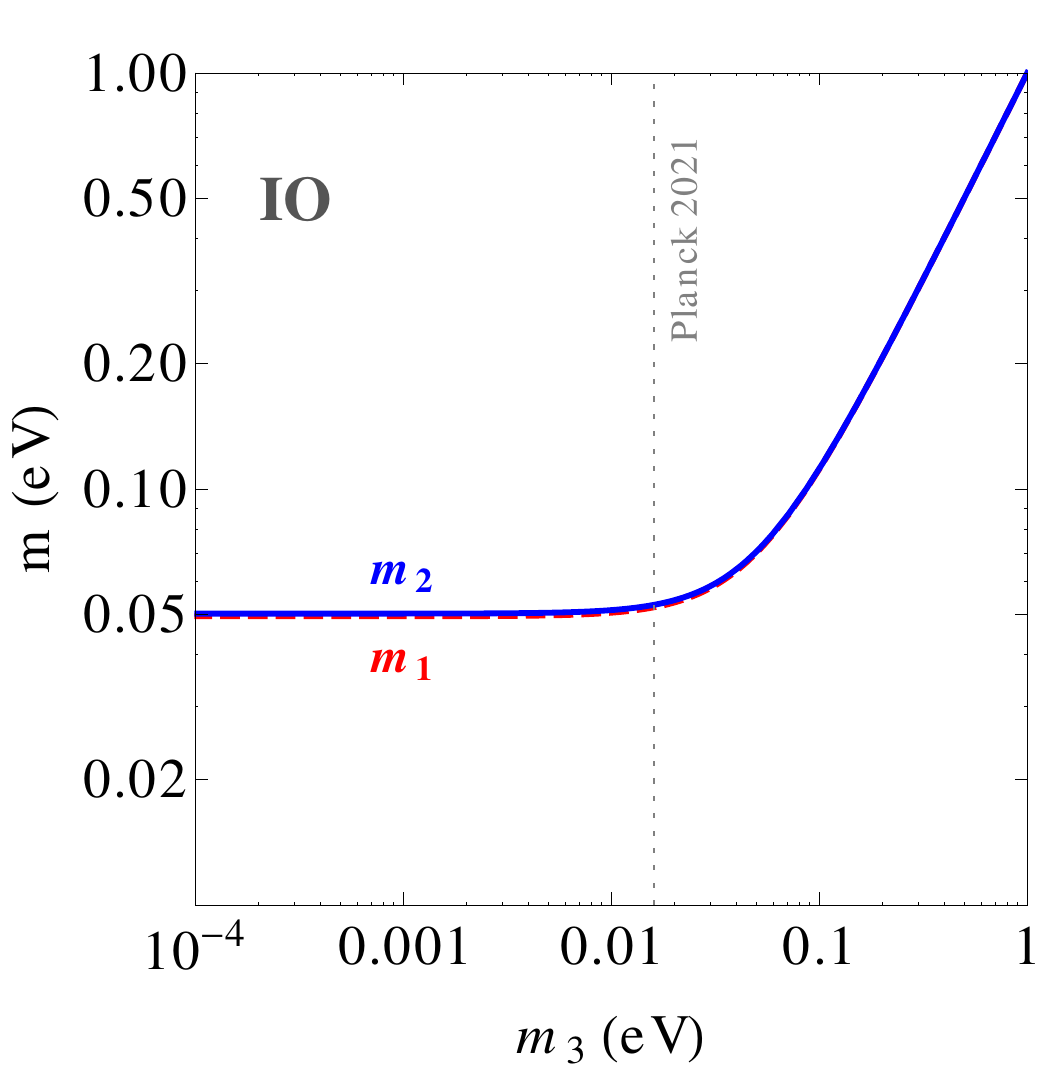}
}%
\caption{Neutrino mass patterns consistent with the results of the global fit, Eqs.~(\ref{mass_fit_2})-(\ref{mass_fit_32}), for the case of (a)~normal ordering~(NO) and (b)~inverted ordering~(IO). The 95\%~C.L. upper bound by Planck, \refeq{planck}, is shown as a gray dashed vertical line.}
\label{fig:Neu_mass}
\end{figure}

The summary of the resulting neutrino mass patterns consistent with Eqs.~(\ref{mass_fit_2})-(\ref{mass_fit_32}), and \refeq{planck} is presented in \reffig{fig:Neu_mass}(a) for NO and in \reffig{fig:Neu_mass}(b) for IO. The degenerate regime, $m_1\approx m_2\approx m_3$, is excluded at the 95\%~C.L. by the Planck measurement (indicated in both panels of \reffig{fig:Neu_mass} by the gray dashed line). At the upper end the largest neutrino masses allowed are 
\bea
\textrm{NO}:&& m_1=0.030\,{\rm eV},\quad m_2=0.031\,{\rm eV},\quad m_3=0.058\,{\rm eV},\label{spl_NO}\\
\textrm{IO}:&& m_1=0.052\,{\rm eV},\quad m_2=0.052\,{\rm eV},\quad m_3=0.016\,{\rm eV}.\label{spl_IO}
\eea

In the hierarchical regime, the lightest neutrino can be much lighter than the other two (actually, it can even be massless), whose masses saturate at 
\bea
\textrm{NO}:&& m_2=0.009\,{\rm eV},\quad m_3=0.050\,{\rm eV},\label{spl_NO_H}\\
\textrm{IO}:&& m_1=0.049\,{\rm eV},\quad m_2=0.050\,{\rm eV}.\label{spl_IO_H}
\eea

In general, there are several known ways to obtain light neutrinos by minimally extending the framework of the SM. The simplest one is to add to the theory three additional Weyl spinors that are singlets under the SM gauge symmetry group, $\nu_{R,i}$, $i=1,2,3$. 
The Yukawa part of the Lagrangian is then extended by 
\be\label{mass:dir}
\mathcal{L}_D=-y_\nu^{ij} \nu_{R,i}\, (H^c)^{\dag} L_j + \textrm{H.c.}\,,
\ee  
where $L_j$, and $H$ are the SM lepton and Higgs boson SU(2)$_L$ doublets, $H^c\equiv i \sigma_2 H^{\ast}$ is the charged conjugate doublet, and a sum over SM generations is implied. The Lagrangian in \refeq{mass:dir}
is renormalizable and does not violate lepton-number symmetry. The left-handed neutrino component of $L_i$ can be combined with a right-handed counterpart, $\nu^{\dag}_{R,i}$, 
to from three Dirac fermions after electroweak-symmetry breaking (EWSB). 
The Dirac mass is generated through the Higgs mechanism as $m_D\sim y_\nu v/\sqrt{2}$, where $v=246\,\textrm{GeV}$ is the Higgs vacuum expectation value (vev). 
To fit the experimental data, the required neutrino Yukawa couplings need to be of the order of $10^{-13}$. For comparison, the Yukawa couplings of the other SM fermions span the range $10^{-5}$ to 1. 

Since the right-handed neutrinos are singlets of the SM gauge group, they can additionally 
acquire a Majorana mass,
\be\label{mass:maj}
\mathcal{L}_M=-\frac{1}{2}M_N^{ij}\nu_{R,i}\,\nu_{R,j}+\textrm{H.c.}\,,
\ee
which violates the conservation of lepton number and can be forbidden if the latter is promoted to being a symmetry of the theory. A possibly more natural alternative to the Dirac-type generation of neutrino masses described above is thus given by 
the see-saw mechanism\cite{Minkowski:1977sc,Gell-Mann:1979vob,Yanagida:1979as,Glashow:1979nm,Mohapatra:1980yp,Schechter:1981cv,Schechter:1980gr}. In its simplest formulation, Type-I,  
 by combining \refeq{mass:dir} and \refeq{mass:maj} one obtains the well-known neutrino mass matrix
\be\label{eq:m_mat}
m_\nu=\left(\begin{array}{cc}
0 & m_D^T \\
m_D & M_N
\end{array}\right)\,.
\ee
If the eigenvalues of the Majorana mass matrix $M_N$ are much larger than typical $m_D$ values, the diagonalization of $m_\nu$ leads to three light Majorana neutrinos with mass $\sim y^2_\nu v^2/(\sqrt{2}M_N)$, and heavy Majorana neutrinos with mass $\sim M_N$. A large Majorana mass $M_N\gg M_t$ can naturally suppress the neutrino mass without the need of considering very small Yukawa couplings in \refeq{mass:dir}. 

Additional known methods for generating a small mass for the neutrinos include radiative generation,
see, \textit{e.g.}, Refs.\cite{Cai:2017jrq,Klein:2019iws} for reviews. 

Partially related to the issue of neutrino masses is that of neutrino mixing\cite{Pontecorvo:1967fh,10.1143/PTP.28.870}. 
For the case of three massive Dirac neutrinos the corresponding PMNS mixing matrix can be conveniently parameterized by three mixing angles and one CP-violating phase as
\begin{eqnarray}
U =\left(
          \begin{array}{ccc}
          c_{12}c_{13} & s_{12}c_{13} & s_{13}e^{-i\delta}  \\
 -c_{23}s_{12} - s_{23}s_{13}c_{12}e^{i\delta} & c_{23}c_{12} -
s_{23}s_{13}s_{12}e^{i\delta}&  s_{23}c_{13}\\
  s_{23}s_{12} - c_{23}s_{13}c_{12}e^{i\delta}& - s_{23}c_{12} -
c_{23}s_{13}s_{12}e^{i\delta} & c_{23}c_{13} \end{array} \right)
\;\;.
\label{PMNS}
\end{eqnarray}
Here $c_{ij} = \cos \theta_{ij}$, $s_{ij} = \sin \theta_{ij}$ and the results of the global fit give at $1\,\sigma$\cite{Gonzalez-Garcia:2012hef, Esteban:2020cvm}
\bea\label{3mixing}
\textrm{NO}:&& 
\theta_{12} = \left(33.44^{+0.77}_{-0.74}\right)^\circ\,, \quad  \theta_{23} = \left(49.2^{+1.0}_{-1.3}\right)^\circ\,, \quad 
\theta_{13} = \left(8.57^{+0.13}_{-0.12}\right)^\circ\,, \quad 
\delta_{\rm{CP}} = \left(194^{+52}_{-25}\right)^\circ\,,\nonumber\\
\textrm{IO}:&& 
\theta_{12} = \left(33.45^{+0.77}_{-0.74}\right)^\circ\,, \quad  \theta_{23} = \left(49.5^{+1.0}_{-1.2}\right)^\circ\,, \quad 
\theta_{13} = \left(8.60^{+0.12}_{-0.12}\right)^\circ\,, \quad 
\delta_{\rm{CP}} = \left(287^{+27}_{-    32}\right)^\circ\,.\nonumber \\
\eea
If, on the other hand, neutrinos are of the Majorana type, two additional phases are present, which remain undetermined by neutrino oscillation experiments. 

Finally, according to the parametrization of the PMNS matrix defined in Eqs.~(\ref{eq:thetas})-(\ref{PMNS2}) of Appendix~\ref{app:appendix}, the allowed $3\,\sigma$ ranges of the mixing angles of \refeq{3mixing} translate into\cite{Esteban:2020cvm} 
\be\label{eq:phen_mix}
X \in [0.64-0.71]\quad Y \in [0.26-0.34] \quad Z \in [0.05-0.26]\quad W \in [0.21-0.48]\,.
\ee

\section{Fixed points in gauge-Yukawa systems \label{sec:fpan}}

\subsection{General notions in AS gravity\label{sec:as}}

Our working assumption is that the SM couples above the Planck scale to quantum gravity or some other NP responsible for generating a fixed point 
for the beta functions of all dimensionless couplings. For the gauge-Yukawa systems investigated in this paper the beta functions of the SM (or NP) are modified in the trans-Planckian regime by parametric corrections, 
\bea\label{eq:gauge-Yuk}
\beta_g&=&\beta_g^{\textrm{SM}}-g\,f_g, \nonumber \\
\beta_y&=&\beta_y^{\textrm{SM}}-y\,f_y, 
\eea
where $\beta_x\equiv dx/d\log k$, $g$ and $y$ are the set of all gauge and Yukawa couplings, respectively, and we   
parameterize with $f_g$ and $f_y$ the effects of new AS trans-Planckian interactions. 
In agreement with theoretical computations from quantum gravity the new corrections are made 
universal, in the sense that they distinguish 
only between different types of matter interactions. Note that in Eqs.~(\ref{eq:gauge-Yuk}) 
we neglect possible effects proportional to higher powers in the matter couplings.

The parameters $f_g$ and $f_y$
should be eventually determined from the gravitational 
dynamics\cite{Zanusso:2009bs,Daum:2009dn,Daum:2010bc,Folkerts:2011jz,Oda:2015sma,Eichhorn:2016esv,Eichhorn:2017eht,Christiansen:2017cxa}.
Specifically, in AS quantum gravity calculations using the functional RG
have shown that $f_g$ is likely to be non-negative, 
irrespective of the chosen RG scheme\cite{Folkerts:2011jz}, 
and that $f_g>0$ is required to enforce asymptotic freedom in the gauge sector. Note that one is naturally led 
to choose an RG scheme in which the leading non-universal coefficient is non-zero to be 
consistent, on the one hand, with the low-energy phenomenology and, on the other, to avoid
having to compute higher-order contributions, which would instead be required to determine the fate of theories with $f_g=0$. 
Interestingly, a non-trivial combined fixed point in a coupled system of gravity and matter was found in Ref.\cite{Christiansen:2017cxa}, where it was proven that gravity can be asymptotically safe, while the gauge sector remains asymptotically free. 

Unlike in the case of $f_g$, the status of the 
leading-order gravitational correction to the Yukawa couplings, $f_y$, is more opaque.
Several simplified models were analyzed in the literature\cite{Rodigast:2009zj,Zanusso:2009bs,Oda:2015sma,Eichhorn:2016esv}, 
but no general results and definite conclusions regarding the size and sign of the leading coefficient are available to our knowledge.

Large uncertainties finally mar determinations of the impact of matter on the gravity sector, which relate to 
the choice of truncation of the gravitational action and, within a chosen truncation, the cutoff-scheme dependence\cite{Reuter:2001ag,Narain:2009qa}.
Early calculations in AS Einstein-Hilbert gravity were performed by retaining two operators in the scale-dependent effective action, with the gravitational dynamics being governed exclusively by the Newton and cosmological constants\cite{Reuter:1996cp}. Inclusion of higher-order 
interactions enriches the theory by additional free parameters\cite{Lauscher:2002sq,Codello:2007bd,Benedetti:2009rx,Falls:2017lst,Falls:2018ylp} and
various results can differ by up to 50-60\%\cite{Dona:2013qba}.

In this work, following the effective approach adopted in some recent articles\cite{Eichhorn:2017ylw,Eichhorn:2018whv,Reichert:2019car,Alkofer:2020vtb,Kowalska:2020gie,Kowalska:2020zve}, we bypass the details and large uncertainties associated with the gravitational dynamics  
and treat $f_g$ and $f_y$ as free parameters that have to be found, eventually, in agreement with the low-scale experimental constraints. Their values thus define a particular set of boundary conditions for the beta functions at the Planck scale. 

A fixed point of the gauge-Yukawa system is given by any set $\{g^\ast,y^\ast\}$, generically denoted with an asterisk,
for which $\beta_g(g^\ast,y^\ast)=\beta_y(g^\ast,y^\ast)=0$.
To determine the structure of the fixed point we
linearize the renormalization group equation (RGE) system of the couplings $\{\alpha_i\}\equiv\{g,y\}$
around the fixed point and derive the stability matrix $M_{ij}$\,:
\be\label{stab}
M_{ij}=\partial\beta_i/\partial\alpha_j|_{\{\alpha^{\ast}_i\}}\,.
\ee
Eigenvalues of $M_{ij}$ define the opposite of the critical exponents $\theta_i$, 
and characterize the power-law evolution of the couplings in the vicinity of $\{\alpha^{\ast}_i\}$. 
If $\theta_i$ is positive the corresponding eigendirection is dubbed as {\it relevant} and UV-attractive. All RG trajectories along this direction will asymptotically reach the fixed point and, as a consequence, a deviation of a relevant coupling from the fixed point introduces a free parameter in the theory. This freedom can be used to adjust the coupling at some high scale to match an eventual measurement in the IR. If $\theta_i$ is negative, 
the corresponding eigendirection is dubbed as {\it irrelevant} and UV-repulsive. 
If $\theta_i<0$ there exist only one trajectory the coupling's flow can follow in its run to the IR, thus providing potentially a clear prediction for its value at the experimentally accessible scale. Finally, $\theta_i=0$ corresponds to a \textit{marginal} eigendirection. The RG flow along this direction is logarithmically slow and one ought to go beyond the linear approximation to decide whether a fixed point is attractive or repulsive.

\subsection{Systems of Yukawa couplings}\label{sec:gen_mech}

As is customary in asymptotic safety we begin with assuming that the interactions of the high-scale theory will 
dynamically solve the triviality problem of the U(1)$_Y$ gauge coupling of the SM. 
For AS gravity, this seems to be in agreement with explicit calculations based on functional RG techniques\cite{Harst:2011zx,Christiansen:2017gtg,Eichhorn:2017lry}.

At one loop in the gauge coupling $g_Y$ the SM RG flow receives a
trans-Planckian correction parameterized by $f_g$\,:
\be\label{eq:gy}
\frac{d g_Y}{dt}=\frac{g_Y^3}{16\pi^2}\,\frac{41}{6}-f_g\, g_Y\,.
\ee
The beta function in \refeq{eq:gy} develops an interactive fixed point. The position of the decoupling scale (Planck) is such to guarantee consistency with the low-energy value of the U(1)$_Y$ coupling. We indicatively select $M_{\textrm{Pl}}=10^{19}\,\textrm{GeV}$, which leads to $g_Y^{\ast}= 0.47$ in the trans-Planckian regime. One thus expects $f_g=(41/6)\, g_Y^{\ast^2}/16 \pi^2=0.0096$.    

It is then reasonable to expect that a realistic theory with asymptotic safety (possibly based on quantum gravity) should not be characterized by large fine tuning of the parameters $f_g$, $f_y$ against the numerical coefficients of the SM RGEs. The former, in fact, should eventually be determined from the calculation of the full quantum-gravity plus matter action. The latter depend instead 
on the gauge quantum numbers and the one-loop structure of the SM Lagrangian, and are independent of quantum gravitational interactions. 
Following these two assumptions (interactive fixed point $g_Y^\ast$ and no fine tuning) we are able to restrict the number of predictive solutions emerging from the fixed-point analysis.      

Let us consider the RGEs of a
system composed of $g_Y$ and two SM (but the discussion can be easily extended to NP) Yukawa couplings indicated generically with $y_X$ and $y_Z$: 
\bea
\frac{d y_X}{d t}&=&\frac{y_X}{16\pi^2}\left[ \alpha_X\, y_X^2+\alpha_Z y_Z^2-\alpha_Y\, g_Y^2\right]-f_y\, y_X\label{eq:yx} \\
\frac{d y_Z}{d t}&=&\frac{y_Z}{16\pi^2}\left[ \alpha'_X\, y_X^2+\alpha'_Z y_Z^2-\alpha'_Y\, g_Y^2\right]-f_y\, y_Z\,,\label{eq:yz}
\eea
where $\alpha^{(\prime)}_X, \alpha^{(\prime)}_Y, \alpha^{(\prime)}_Z > 0$ are coefficients of order one. 
While there are strong indications from explicit calculations
that $f_g>0$\cite{Folkerts:2011jz}, 
no equivalent precision has been yet achieved for $f_y$.  

Since we seek to maximize the predictivity of the system at the low scale we focus on fixed-point solutions of two kinds. Either the fixed point is fully interactive, $y^{\ast}_X\neq 0$,  $y^{\ast}_Z\neq 0$, or one of the Yukawa couplings is interactive 
at the fixed-point and the other is Gaussian: $y^{\ast}_X\neq 0$,  $y^{\ast}_Z= 0$.
In the case of a fully interactive fixed point one gets
\bea
y_X^{\ast 2}&=& \left(16 \pi^2 f_y + \alpha_Y g_Y^{\ast 2}-\alpha_Z y_Z^{\ast 2}\right)/\alpha_X  \label{eq:ft_a} \\
y_Z^{\ast 2}&=& \left(16 \pi^2 f_y + \alpha_Y' g_Y^{\ast 2}-\alpha'_X y_X^{\ast 2}\right)/\alpha'_Z\,. \label{eq:ft_b}
\eea
Barring accidental cancellations, for a generic $f_y$ assumed to be reasonably 
of the same order of magnitude as 
$f_g$ ($|f_y| \approx 10^{-4}-10^{-2}$), no fine tuning between the addends of the right-hand sides of Eqs.~(\ref{eq:ft_a}), (\ref{eq:ft_b})
implies that $y_X^{\ast}$, $y_Z^{\ast}$ should be themselves approximately of the size
of $g_Y^{\ast}$. If we restrict ourselves to the couplings of the SM, this is in agreement with observations at the low scale for the top quark only. Therefore, it is reasonable 
to use the top quark Yukawa coupling to fix (or ``guess'') the value of $f_y$, as all other Yukawa couplings, being much smaller, would require some level of fine tuning of $f_y$ against $g_Y^{\ast 2}$. In the phenomenological approach,
one will invert \refeq{eq:ft_a} to derive the ``window'' of $f_y$ values consistent with a trans-Planckian fixed point 
$y_X^{\ast}\equiv y_t^{\ast}$ in agreement with the low-scale determination, $0.94\leq y_t(M_t)\leq 1$\cite{Eichhorn:2018whv}. Encouragingly, this approach leads to numerical values of $f_y$ that are 
highly consistent with the results of explicit calculations based on the functional RG\cite{Eichhorn:2017ylw}, especially given the large theoretical uncertainties\cite{Dona:2013qba}.
Note, however, that once we 
use $y_X^{\ast}\equiv y_t^{\ast}$ to extract the numerical value of $f_y$ 
any other fixed-point Yukawa coupling $y_Z^{\ast}\neq 0$ is expected to be roughly of the same size as $y_X^{\ast}$.

Let us consider then the second type of fixed point, featuring one of the Yukawa couplings as interactive and the other Gaussian: $y^{\ast}_X\neq 0$ given by \refeq{eq:ft_a}, and $y^{\ast}_Z= 0$. Whether $y^{\ast}_Z= 0$ corresponds to a relevant or irrelevant direction depends now on the interplay of several factors:  the hypercharge assignments of the Weyl components of fermion $Z$ -- not to be confused with the neutrino mixing parameter $Z$ in \refeq{eq:phen_mix} -- which determines 
the (positive) coefficient $\alpha_Y'$; its one-loop Yukawa coefficient $\alpha_X'$; the size of fixed points $y_X^{\ast}$, $g_Y^{\ast}$; and the value of $f_y$. 
In fact, in the approximation where the stability matrix is diagonal,
one can express the critical exponent of $y_Z^{\ast}=0$  as
\be\label{eq:irr}
\theta_Z\approx -\left(\alpha_X'\, y_X^{\ast 2}  -\alpha_Y'\,g_Y^{\ast 2}- 16 \pi^2 f_y\right)\,.
\ee
By inserting \refeq{eq:ft_a} into \refeq{eq:irr} one can determine the critical value of $f_y$ which separates two regions of distinct scaling 
behavior of coupling $y_Z$ in the vicinity of its Gaussian fixed point:
\be\label{eq:crit_fy}
f_{Z, X Y}^{\textrm{crit}}=\frac{g_Y^{\ast 2}}{16 \pi^2}\,\frac{\alpha_X' \alpha_Y-\alpha_Y' \alpha_X}{\alpha_X-\alpha_X'}\,.
\ee
If $f_y >f_{Z, X Y}^{\textrm{crit}}$ then $y_Z^{\ast}=0$ along a relevant, UV-attractive direction. If $f_y <f_{Z, X Y}^{\textrm{crit}}$ then $y_Z^{\ast}=0$ is irrelevant or IR-attractive. 
One can easily integrate \refeq{eq:yz} to obtain 
the flow of $y_Z$. In terms of an arbitrary constant 
$\kappa$, which can be used to set the position of the Planck scale along the flow, one finds
\be\label{eq:analytic}
y_Z(t,\kappa)=\left[\frac{16 \pi^2 c_X \left(f_{Z, X Y}^{\textrm{crit}}-f_y \right)}{e^{2 c_X \left(f_{Z, X Y}^{\textrm{crit}}-f_y\right)\left(16\pi^2 \kappa-t\right)}+\alpha_Z'}\right]^{1/2}\,,
\ee
where $c_X\equiv (\alpha_X-\alpha_X')/\alpha_X$.

\setlength\tabcolsep{0.25cm}
\begin{table}[t]\footnotesize
\begin{center}
\begin{tabular}{|c|c|c|c|}
\hline
$Z$ & $\alpha_{X=t}'$ & $\alpha_Y'$ & $ f_{Z, t Y}^{\textrm{crit}}$ \\
\hline
$u,c$ & $3$ & $ \frac{17}{12}$ & $-20.0\times 10^{-4}$ \\
\hline
$b$ & $\frac{3}{2}$ & $ \frac{5}{12}$ & $1.17\times 10^{-4}$ \\
\hline
$d,s$ & $3$ & $ \frac{5}{12}$ & $22.3\times 10^{-4}$ \\
\hline
$\nu_i$ & $3$ & $ \frac{3}{4}$ & $8.22\times 10^{-4}$ \\
\hline
$e,\mu,\tau$ & $3$ & $ \frac{15}{4}$ & $-119 \times 10^{-4}$ \\
\hline
\end{tabular}
\caption{The values of $\alpha_{X=t}'$, $\alpha_Y'$, and $f_{Z, t Y}^{\textrm{crit}}$ for the Yukawa coupling of fermion $Z$.}
\label{tab:critSM}
\end{center}
\end{table}

For illustration, let us assume that $X=t$, so that $\alpha_X=9/2$ and $\alpha_Y=17/12$. The values of $\alpha_X'$, $\alpha_Y'$, and $f_{Z, X Y}^{\textrm{crit}}$ for the remaining Yukawa couplings $y_Z$ in the SM are listed in \reftable{tab:critSM}. We note that $y_{u,c,e, \mu,\tau}^{\ast}=0$ 
along a relevant direction, unless $f_y$ is very large and negative (a scenario that tends to be in tension with the determination of the top mass). 
Conversely, when $Z=b,d,s,\nu_i$, $y_Z^{\ast}=0$ is an IR-attractive fixed point for moderate values of $f_y$.

Finally, two distinct regions of interest, corresponding to $f_y$ larger or smaller than the critical value, 
also emerge in the case 
of the fully interactive fixed point $y^{\ast}_{X}\neq 0$,  $y^{\ast}_Z\neq 0$ 
described in Eqs.~(\ref{eq:ft_a}), (\ref{eq:ft_b}). 
While for $f_y > f_{Z, X Y}^{\textrm{crit}}$ the fully interactive IR-attractive solution encoded in the equations yields real Yukawa couplings,  
for $f_y < f_{Z, X Y}^{\textrm{crit}}$ one of the Yukawa couplings becomes imaginary.

\subsection{The top/neutrino case\label{sec:top-neu}}

We now further restrict the observations of \refsec{sec:gen_mech} to the specific case of the SM top quark and neutrino Yukawa couplings: 
$y_X\equiv y_{t}$, $y_Z\equiv y_{\nu}$. In fact, the IR-attractive solution at $y_{\nu}^{\ast}=0$
would provide a rationale for expecting at the low scale 
a minuscule neutrino Yukawa coupling without fine tuning. 
We implicitly 
assume that all SM Yukawa couplings with the exception of the top quark and neutrino
are set at relevant Gaussian fixed points (this requires some care, see Footnote~\ref{foot:ds}).  
The trans-Planckian RGEs are thus simplified as [cf.~Eqs.~(\ref{eq:gy})-(\ref{eq:yz})]
\begin{eqnarray}
\frac{d g_Y}{dt}&=&\frac{g_Y^3}{16\pi^2}\,\frac{41}{6}-f_g\, g_Y\label{eq:betagy}  \\
\frac{d y_t}{dt}&=&\frac{y_t}{16\pi^2}\left[ \frac{9}{2}y_t^2+y^2_\nu-\frac{17}{12}g_Y^2\right]-f_y\, y_t \label{eq:betayt} \\
\frac{d y_{\nu}}{d t}&=&\frac{y_{\nu}}{16\pi^2}\left[ 3 y_t^2+\frac{5}{2}y_{\nu}^2-\frac{3}{4}g_Y^2\right]-f_y\, y_{\nu}\,. \label{eq:betay1}
\end{eqnarray}

As was mentioned in \refsec{sec:gen_mech},
we deem acceptable phenomenological agreement with the top mass at the low scale if $0.94\leq y_t(M_t)\leq 1$. 
Equation~(\ref{eq:betayt}) will thus induce a phenomenologically allowed window, $-0.0001\lesssim f_y\lesssim 0.001$. 
Following \refsec{sec:gen_mech}
we next proceed to identify two scenarios, 
depending on whether $f_y$ is larger or smaller than the critical value, 
$f_{\nu,t Y}^{\textrm{crit}}\approx 8.2 \times 10^{-4}$. If $f_y > f_{\nu,t Y}^{\textrm{crit}}$ the system admits a maximally predictive IR-attractive fixed point, interactive in all three of the couplings in Eqs.~(\ref{eq:betagy})-(\ref{eq:betay1}). 
In this scenario $y_\nu^{\ast}$ is real, IR-attractive, and not much smaller than $g_Y^{\ast}$ and $y_t^{\ast}$. Therefore, it cannot give rise to a neutrino mass of the Dirac type.

One might assume for this scenario the presence of a type-I see-saw mechanism at some scale 
below the Planck scale, cf.~the discussion surrounding Eqs.~(\ref{mass:maj}), (\ref{eq:m_mat}) in \refsec{sec:neutrino}. In order to fit the neutrino spectrum one may associate, phenomenologically, one value of the Majorana mass-scale $M_N$ to every emerging $y_{\nu}(M_N)$ as function of $f_y$.\footnote{The impact of AS quantum gravity calculations on the RGEs of the Majorana mass term was 
investigated in detail, \textit{e.g.}, in Refs.\cite{DeBrito:2019rrh,Hamada:2020vnf,Domenech:2020yjf}.} 
An upper bound exists, imposed by the need to fit the top mass at the EWSB scale with the same $f_y$. For $f_y\lesssim 0.001$ one gets
$M_N\lesssim 1.4 \times 10^{12}\,\textrm{GeV}$. 
Conversely, arbitrarily small values of the Majorana mass are theoretically allowed in this scenario, but at the price of increasing drastically the fine tuning of the r.h.s.~of \refeq{eq:betay1}. 

 \begin{figure*}[t]
	\centering%
		\includegraphics[width=0.44\textwidth]{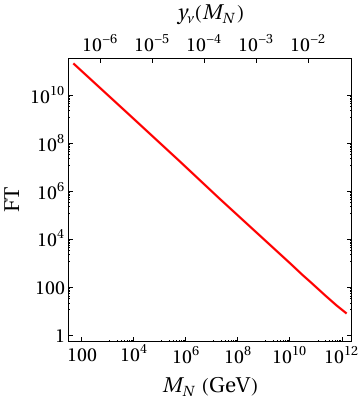}
\caption{For the case $f_y >f_{\nu,t Y}^{\textrm{crit}}$, the fine tuning (FT) of the parameter $f_y$ corresponding to different choices of the Majorana mass and neutrino Yukawa coupling in Type-I see-saw scenarios.}
\label{fig:MassMaj}
\end{figure*}

One can define a fine tuning measure, 
\be
\textrm{FT}=\frac{16 \pi^2 f_y}{5/2\, y_{\nu}^{\ast 2}}\,.
\ee
In \reffig{fig:MassMaj} 
we plot the fine tuning of $f_y$ versus the expected Majorana mass and the neutrino 
Yukawa coupling in type-I see-saw scenarios. 

\setlength\tabcolsep{0.25cm}
\begin{table}[t]\footnotesize
\begin{center}
\begin{tabular}{|c|ccc|ccc|}
\hline
 & $g_Y^\ast$ & $y_t^\ast$ & $y_\nu^\ast$ & $\theta_Y$ & $\theta_t$ & $\theta_\nu$ \\
\hline
FP$_{\textrm {A}}$ & $ 0$ & $ 0$ & $0$ & $+$ & $+$ & $+$ \\
FP$_{\textrm {B}}$ & $0$ & $0$ & $\ast$ & $+$ & $+$ & $-$ \\
FP$_{\textrm {C}}$ & $0$ & $\ast$ & $0$ & $+$ & $-$ & $+$ \\
FP$_{\textrm {D}}$ & $0$ & $\ast$ & $\ast$ & $+$ & $-$ & $-$ \\
FP$_{\textrm {E}}$ & $\ast$ & $0$ & $0 $ & $-$ & $+$ & $+$ \\
FP$_{\textrm {F}}$ & $\ast$ & $\ast$ & $0$ & $-$ & $-$ & $-$ \\
FP$_{\textrm {G}}$ & $\ast$ & $0$ & $\ast$ & $-$ & $+$ & $-$ \\
\hline
\end{tabular}
\caption{The real fixed points of the system of Eqs.~(\ref{eq:betagy})-(\ref{eq:betay1}) when $f_y<f_{\nu,t Y}^{\textrm{crit}}$. The asterisk indicates a non-zero value. In the right-hand side box the signs of the relative critical exponents are given.}
\label{tab:TM_FP}
\end{center}
\end{table}

The second region of interest arises instead for $f_y <  f_{\nu,t Y}^{\textrm{crit}}$. The system of Eqs.~(\ref{eq:betagy})-(\ref{eq:betay1}) admits several fixed points, 
which are listed in \reftable{tab:TM_FP} with the sign of their respective critical exponents. 
Fixed point FP$_\textrm{F}$ features the IR-attractive solution with $y_{\nu}^{\ast}=0$ discussed in generic terms in \refsec{sec:gen_mech}. (A non-interactive irrelevant fixed-point 
direction for the neutrino Yukawa coupling
was identified also in Ref.\cite{Held}. The dependence 
of the critical exponents of the SM lepton and neutrino Yukawa couplings on the hypercharge assignment was there explored.) 

\begin{figure}[t]
\centering
\subfloat[]{%
\includegraphics[width=0.44\textwidth]{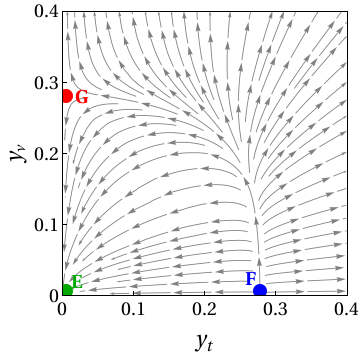}
}%
\hspace{0.02\textwidth}
\subfloat[]{%
\includegraphics[width=0.45\textwidth]{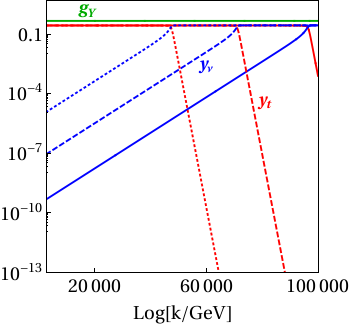}
}%
\caption{(a) Phase diagram in the plane ($y_t$, $y_\nu$) for the three fixed points featuring $g_Y^\ast\neq 0$ in \reftable{tab:TM_FP}. The RG flow directions point towards the UV.  (b) The flow of the top (red) and neutrino (blue) Yukawa coupling for three different choices (solid, dashed, dotted) 
of the $y_t$ trajectory joining the UV fixed point FP$_\textrm{G}$ to the IR fixed point FP$_\textrm{F}$. The abelian gauge coupling is shown in solid green. In both panels the AS gravity parameters are set to $f_g=0.0096$ and $f_y=0.0002$. }
\label{fig:TM_Phase}
\end{figure}

While an IR-attractive direction with $y_\nu^{\ast}=0$ provides the central 
ingredient for generating dynamically an infinitesimally small coupling, the self-standing FP$_\textrm{F}$ 
predicts zero neutrino Yukawa couplings at all scales, in contrast with observations. 
It is then crucial to point out that all other solutions in \reftable{tab:TM_FP}, which are UV-attractive, 
eventually fall into the basin of attraction of FP$_\textrm{F}$. This can be seen explicitly in the phase diagram presented in \reffig{fig:TM_Phase}(a), where we highlight the three fixed points corresponding to an interactive irrelevant direction for $g_Y$. As the arrows point toward the UV, one can see that the IR-attractive fixed point FP$_\textrm{F}$
can be reached from any other UV fixed point along the flow towards the low energy. Note that similar trajectories, not shown in the picture, merge the flow of the four remaining fixed points corresponding to $g_Y^{\ast}=0$ into FP$_\textrm{F}$. This happens as soon as $g_Y$ deviates from the zero critical surface in its running to the IR.  

If the Planck scale is met before the flow merges completely into FP$_\textrm{F}$ one observes the decoupling of trans-Planckian interactions 
when the Yukawa coupling in the neutrino sector is arbitrarily small. This solution is thus generic and UV-complete. The only parameter regulating the size of the neutrino Yukawa couplings is the trajectory followed by the top Yukawa coupling along the flow or, equivalently, the position of the Planck scale along the flow, parameterized by $\kappa$ in \refeq{eq:analytic}. 
Whether Nature has selected one or the other trajectory is eventually a matter of experimental determination, 
and it is not subject to fine tuning. 

In \reffig{fig:TM_Phase}(b) we show the flow of the top Yukawa coupling (in red), and the neutrino Yukawa coupling (in blue) for three different parametrizations (solid, dashed, dotted) of the $y_t$ trajectory joining the UV fixed point FP$_\textrm{G}$ to the IR fixed point FP$_\textrm{F}$. The chosen trajectory determines the final value of $y_\nu$ at the Planck scale.  Note the extremely high value of the UV scale, which is due to the slowness of the neutrino Yukawa running. A similar behavior was observed in Ref.\cite{Alkofer:2020vtb} for the CKM matrix elements. Despite being an unusual feature, it does not constitute a problem since the flow remains finite up to infinity.\footnote{\label{foot:ds}The attentive reader may have noticed that our underlying assumption of having all but the top quark and neutrino Yukawa couplings set at Gaussian relevant fixed points in Eqs.~(\ref{eq:betagy})-(\ref{eq:betay1}) 
cannot be fulfilled for $y_d$ and $y_s$, as the upper bound on $f_y\lesssim 0.001$ lies below $f_{d(s), t Y}^{\textrm{crit}}\approx 0.002$ given in \reftable{tab:critSM}. It follows that for $y_t^{\ast}\neq 0$, 
$y_{d(s)}^{\ast}= 0$ would be IR attractive. In fact, in the absence of CKM mixing, for the $f_y$ values selected here $y_d$ and $y_s$ would behave in similar fashion to the right-handed neutrino Yukawa couplings 
and thus produce a phase diagram akin to \reffig{fig:TM_Phase}(a). However, it was pointed out in Ref.\cite{Alkofer:2020vtb} 
that introducing CKM mixing induces a modification in the RGEs large enough to render $y_{d(s)}$ IR-repulsive. 
As a consequence, their trajectory can be matched freely to the low-scale measured value of the corresponding SM fermion.}

 \begin{figure*}[t]
	\centering%
	\includegraphics[width=0.7\textwidth]{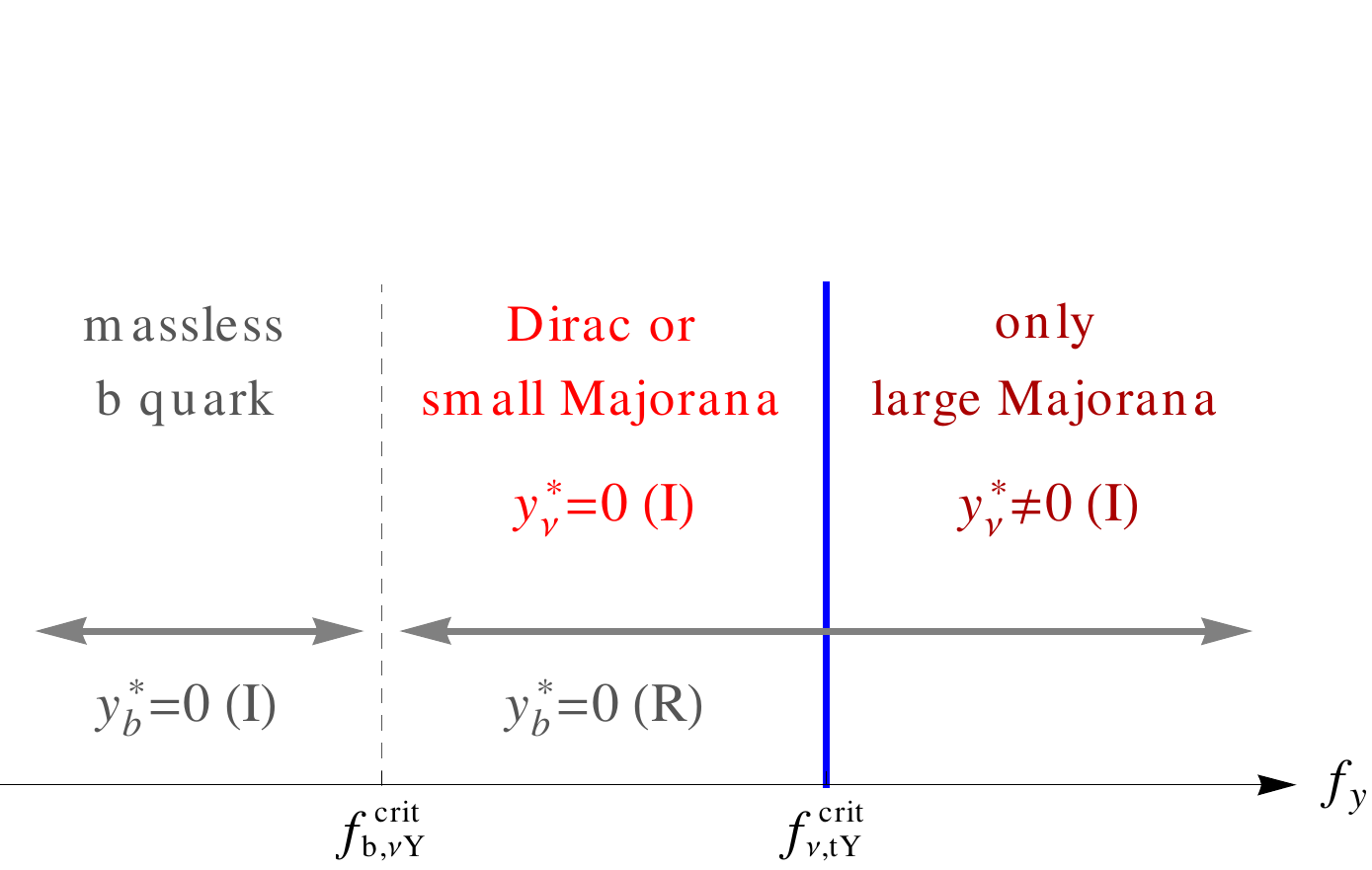}
\caption{Implications for the low-energy 
particle spectrum as a function of the gravitational parameter $f_y$. In the region $f_y > f_{\nu,t Y}^{\textrm{crit}}$, 
where $f_{\nu,t Y}^{\textrm{crit}}\approx 8.2 \times 10^{-4}$ in the SM plus right-handed neutrinos, fixed-point
value $y_{\nu}^{\ast}$ (irrelevant) is of order one. One requires a large Majorana mass in the Lagrangian to suppress the SM neutrino mass. In the region $f_{b,\nu Y}^{\textrm{crit}}< f_y < f_{\nu,t Y}^{\textrm{crit}}$ one gets $y_{\nu}^{\ast}=0$ along an irrelevant direction of the flow: neutrinos of the Dirac type can naturally get their mass via the Higgs mechanism, or one might introduce light sterile neutrinos. In the region $f_y <f_{b,\nu Y}^{\textrm{crit}}$, where the value $f_{b,\nu Y}^{\textrm{crit}}$ depends on the selected phase-space trajectory (see main text), 
the bottom Yukawa coupling features a Gaussian fixed point along an irrelevant direction of the flow, which leads to the prediction of a massless bottom quark.}
\label{fig:RangeF}
\end{figure*}

We summarize the main properties of the particle spectrum in different regions of the $f_y$ parameter space in \reffig{fig:RangeF}.
In a realistic setting, one would have to include modifications to Eqs.~(\ref{eq:betagy})-(\ref{eq:betay1}) due to the second most sizable Yukawa coupling, \textit{i.e.}, the bottom quark's.    
The lower bound $f_{b,\nu Y}^{\textrm{crit}}$ is placed here to indicate where
one gets $y_b^{\ast}=0$ along a relevant, rather than irrelevant, direction.

As can be inferred from \refeq{eq:ft_b} with $X=t$, $Z=b$, for $f_y\gsim f_{b,t Y}^{\textrm{crit}}=1.17\times 10^{-4}$ (\reftable{tab:critSM}) 
there exists an IR-attractive fixed point with real non-zero values for both $y_t^{\ast}\neq 0$, $y_b^{\ast}\neq 0$. Generally, the resulting $y_b^{\ast}$ tends to be too large to predict the correct mass of the bottom quark at the EWSB scale. In principle, however, the correct bottom Yukawa coupling can still be made to emerge, at the price of a mild fine tuning of $f_y$ against the other addends on the right-hand side of \refeq{eq:ft_b}. In other words, $f_y$ should approach closely from above the critical value 
$f_{b,t Y}^{\textrm{crit}}$. In order to get $y_b(M_t)=0.018$, one needs a tuning of approximately one part in $10^2$. This is the fixed-point solution favored in Refs.\cite{Eichhorn:2018whv,Held}, which has the advantage, with respect to solutions where $y_b^{\ast}=0$ is relevant, of correctly predicting the ratio $M_t/M_b$ directly from the UV completion. However, in the presence of UV-attractive fixed points like
FP$_\textrm{E}$ and FP$_\textrm{G}$ in \reftable{tab:TM_FP},
it is $f_{b,\nu Y}^{\textrm{crit}}$ that provides the lower bound below which 
$y_b^{\ast}=0$ becomes irrelevant. Equation~(\ref{eq:crit_fy}) can be used to compute $f_{b,\nu Y}^{\textrm{crit}}$ at the fixed point. One finds $f_{b,\nu Y}^{\textrm{crit}}\approx -5.8\times 10^{-4}$ at 
FP$_\textrm{E}$ and  $f_{b,\nu Y}^{\textrm{crit}}\approx 1.8\times 10^{-4}$ at 
FP$_\textrm{G}$.

\section{Fixed points in the SM with right-handed neutrinos\label{sec:SM}}

We can extend now 
the discussion of \refsec{sec:gen_mech} to the leptonic sector of SM extended by 3 right-handed neutrinos, $\nu_{R,i}$ with $i=1,2,3$. 
We use the complete set of trans-Planckian one-loop RGEs in the mass basis, which include the neutrino Yukawa couplings, charged-lepton Yukawa couplings, and neutrino mixing angles. The full set of RGEs is presented in Appendix~\ref{app:appendix}. 

We restrict our analysis to the case of Dirac neutrinos, 
for which the trans-Planckian generation of small couplings finds its best justification.\footnote{While it is not currently known empirically if neutrinos are Dirac or Majorana, we expect that tiny Yukawa couplings, 
associated with Dirac or pseudo-Dirac neutrinos, lead to a specific phenomenology. In particular, (pseudo-)Dirac neutrinos are characterized by a tiny magnetic moment\cite{Fujikawa:1980yx,Pal:1981rm};
if neutrinos are Dirac,
the neutrinoless double beta decay ($0\nu2\beta$) process\cite{PhysRev.56.1184,Rodejohann:2012xd,Bilenky:2014uka,DellOro:2016tmg} will not be observed in ongoing experiments;
a tiny neutrino Yukawa coupling does not leave an imprint on cosmological observables like the effective number of neutrino species $N_{\textrm{eff}}$\cite{PhysRevLett.45.669,PhysRevD.25.1481}. Intriguingly, however, it may be possible to distinguish Dirac and pseudo-Dirac neutrinos in neutrino-oscillation experiments on cosmic scales\cite{Beacom:2003eu} and potentially, for certain regions of the parameter space, in $0\nu2\beta$\cite{Lindner:2014oea}.}  
As was discussed in \refsec{sec:top-neu}, this implies   
$f_y<f_{\nu, t Y}^{\textrm{crit}}$ to avoid fine tuning in the gravitational parameters. 

We expect 
the emergence of two fixed-point solutions: a fully IR-attractive fixed point characterized by 
$y_t^{\ast}\neq 0$, $y_{\nu i}^{\ast}=0$, and a
UV-attractive fixed point with a relevant top Yukawa and an irrelevant non-zero neutrino Yukawa coupling,
\be\label{eq:UVfp}
y_t^{\ast}= 0\,(\textrm{R})\quad \quad y_{\nu i}^{\ast}\neq 0\,(\textrm{I}).
\ee
All other Yukawa couplings are set at a Gaussian UV-attractive fixed point.

Since gravitational interactions are flavor-blind by construction, and all charged-lepton Yukawa couplings are set to zero at the fixed point to be in agreement with their SM value, all interactive fixed points of the neutrino Yukawa couplings are degenerate. This fact allows one to investigate different UV scenarios: three degenerate interactive neutrino Yukawa couplings; two irrelevant  
degenerate couplings and one Gaussian relevant Yukawa direction; one irrelevant interactive coupling and two Gaussian relevant directions.

We do not linger here on the first two cases for several reasons. 
To begin with, the fully degenerate interactive case is excluded for Dirac neutrinos on phenomenological grounds. 
As Eqs.~(\ref{eq:y1rge})-(\ref{eq:y3rge}) in Appendix~\ref{app:appendix} show, 
the splitting of different neutrino generations is proportional to the running of the charged lepton Yukawa couplings.   
As the latter acquire a significant non-zero value very late in their flow to the low-scale, an RG-induced splitting
of the third generation over the other two reads $|\Delta m_{\nu}/m_{\nu}|\approx 10^{-4}$ at the EWSB scale, 
which is not in line with observations of the cosmological bound\cite{Planck:2018vyg}, see also Eqs.~(\ref{spl_NO}), (\ref{spl_IO}).\footnote{While for Dirac neutrinos a (nearly) degenerate UV fixed point appears to be in tension with observations at the low scale, the case may be different for Majorana neutrinos, see, \textit{e.g.}, Refs.\cite{Chankowski:1999xc,Antusch:2005gp} for the corresponding RGEs. The experimental absence of neutrino oscillations beyond the known solar and atmospheric frequencies imposes a lower bound on the Majorana mass, $M_N\gsim 1\,\textrm{eV}$\cite{deGouvea:2005er,Donini:2011jh}. One finds,
on purely dimensional grounds, that three sterile neutrino masses as light as $1-2\,\textrm{eV}$
can induce mass splitting of the right order when $y_{\nu,i}(M_t)\approx 1.4\times  10^{-12}$\,.}

More importantly, it has long been known\cite{Chankowski:1999xc,Lindner:2005as} that 
the beta functions of the mixing angles are subject to instability when the neutrino Yukawa couplings become very degenerate, 
so that these cases 
may be inconsistent with a UV completion based on asymptotic safety. 
As can be seen in Eqs.~(\ref{eq:Xrge})-(\ref{eq:Wrge}) of Appendix~\ref{app:appendix}, at a fixed point characterized by Gaussian charged leptons and at least two degenerate neutrino Yukawa couplings,  
a uniquely defined limit for the beta functions of $X$, $Y$, $Z$, and $W$ does not seem to exist, as by  
approaching $t\to \infty$ from different directions in theory space one obtains different 
results.\footnote{One particular solution 
may rely, for example,
on the fact that the size of the splitting of two nearly degenerate neutrino Yukawa couplings along the 
separatrix connecting the UV-attractive interactive fixed point to the IR-attractive Gaussian fixed point can be 
parameterized with the flow of one or more (relevant) charged-lepton Yukawa couplings. If ratios like $y_{\tau}^2[t]/(y_{\nu 1}^2[t]-y_{\nu 2}^2[t])$ in \refeq{eq:Xrge} remain finite for the entirety of the trans-Planckian flow (while $y_e^2[t]$, $y_{\mu}^2[t]$ are hierarchically suppressed) 
there may exist UV fixed points for the mixing angles, of the type
$X^{\ast}=1$, $Y^{\ast}=0$,  $Z^{\ast}=0$,  $W^{\ast}=0$ and others that, when plugged
into \refeq{PMNS2}, complete a faithful  
representations of the permutation group of three objects 
(see Ref.\cite{Alkofer:2020vtb} for the corresponding discussion in the quark case).
However, even if one were able to prove that fixed points of this class represent well-defined solutions for the full system of beta functions, 
all the points that we have tested in this scenario seem to feature an IR-attractive direction of value zero or one for at least one of the mixing parameters, and as such they are excluded, again, on phenomenological grounds.}    

In light of these obstacles we focus here 
on the scenario where only one of the neutrino generations stems from an irrelevant interactive direction in the deep UV, reminiscent of FP$_{\textrm{G}}$ in \refsec{sec:top-neu},  
while the other two come from relevant fixed points, 
reminiscent of FP$_{\textrm{E}}$ in \refsec{sec:top-neu}. 
Because there exists an 
IR-attractive $y_{\nu i}^{\ast}=0$ fixed point for all three generations, 
reminiscent of FP$_{\textrm{F}}$ in \refsec{sec:top-neu}, 
the hierarchy of neutrino Yukawa couplings 
set in the deep UV is conserved along the entirety of the RG flow down to the IR. 

For NO in the neutrino masses, we choose $y_{\nu 3}^{\ast}\neq 0$. 
The fixed points of the system~(\ref{eq:Xrge})-(\ref{eq:Wrge}) are presented in \reftable{tab:SM_FP}, together with the sign of their critical exponents.  
Nine different viable solutions emerge. Two of them correspond to fully irrelevant fixed points attracting all the trajectories in the trans-Planckian IR. They are denoted as FP$_{\textrm {IR,1}}$ and FP$_{\textrm {IR,2}}$ in  \reftable{tab:SM_FP}. The remaining fixed points are UV-attractive and one can show that trajectories stemming from FP$_{\textrm{3,7}}$ end up falling into the basin of attraction of FP$_{\textrm {IR,2}}$, whereas those out of FP$_{\textrm{4-6,8,9}}$ end up falling into the basin of attraction of FP$_{\textrm {IR,1}}$.
The value of the mixing parameters are expected to reproduce \refeq{eq:phen_mix} at the low scale. 
We are able to track the flow down to the correct values for one single set of solutions, 
which conjoin the fully UV-attractive fixed point FP$_5$ with FP$_{\textrm {IR,1}}$. Note, however, that  
FP$_{\textrm {IR,1}}$ is never fully reached, as the flow of the system is drastically modified in the trans-Planckian IR 
by the decreasing size of the $y_{\nu 3}$ Yukawa coupling. 

\setlength\tabcolsep{0.25cm}
\begin{table}[t]\footnotesize
\begin{center}
\begin{tabular}{|c|cccc|cccc|}
\hline
 & $X^\ast$ & $Y^\ast$ & $Z^\ast$ & $W^\ast$ & $\theta_X$ & $\theta_Y$ & $\theta_Z$ & $\theta_W$ \\
\hline
FP$_{\textrm {IR,1}}$ & $1-Y^\ast$ & \rm{ind.} & $Y^\ast$ & $1-Y^\ast$ & $0$ & $-$ & $-$ & $-$\\
FP$_{\textrm {IR,2}}$ & $0$ & $1$ & $1$ & $0$ & $-$ & $-$ & $-$ & $-$ \\
\hline
FP$_{\textrm {3}}$ & $0$ & $1$ & \rm{ind.} & $0$ & $-$ & $-$ & $+$ & $0$ \\
FP$_{\textrm {4}}$ & $0$ & $0$ & \rm{ind.}& $1$ & $+$ & $+$ & $+$ & $-$ \\
FP$_{\textrm {5}}$ & $0$ & $0$ & $1-W^\ast$ & \rm{ind.} & $+$ & $+$ & $+$ & $0$ \\
FP$_{\textrm {6}}$ & $0$ & $0$ & $1$ & $0$ & $+$ & $+$ & $-$ & $+$ \\
FP$_{\textrm {7}}$ & $0$ & $1$ & $0$ & $0$ & $-$ & $0$ & $+$ & $0$ \\
FP$_{\textrm {8}}$ & $0$ & $0$ & $0$ & $1$ & $+$ & $+$ & $+$ & $-$ \\
FP$_{\textrm {9}}$ & $0$ & $0$ & $1$ & $1$ & $+$ & $+$ & $-$ & $-$ \\
\hline
\end{tabular}
\caption{The fixed points of the system~(\ref{eq:Xrge})-(\ref{eq:Wrge}) with $y_{\nu 3}^{\ast}\neq 0$ and all the other lepton Yukawa couplings set at zero. In the right-hand side box the signs of the relative critical exponents are given. The abbreviation ``ind.'' indicates an indeterminate value of the corresponding mixing parameter.}
\label{tab:SM_FP}
\end{center}
\end{table}

\begin{figure}[t]
\centering
\subfloat[]{%
\includegraphics[width=0.44\textwidth]{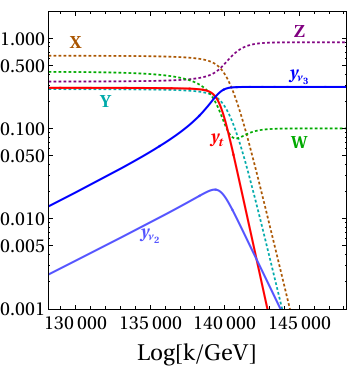}
}%
\hspace{0.02\textwidth}
\subfloat[]{%
\includegraphics[width=0.44\textwidth]{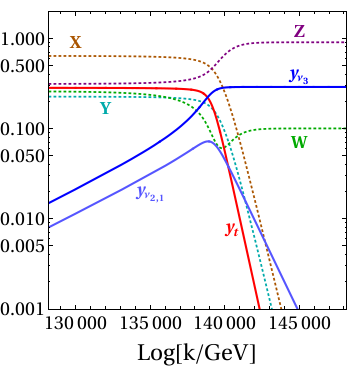}
}%
\caption{(a) RGE trajectories stemming from FP$_5$ of \reftable{tab:SM_FP}, producing in the IR
the neutrino Yukawa couplings given in \refeq{spl_NO_H} and mixing parameters in agreement with \refeq{eq:phen_mix}. (b) RGE trajectories stemming from FP$_5$, producing in the IR
the neutrino Yukawa couplings given in \refeq{spl_NO} and mixing parameters in agreement with \refeq{eq:phen_mix}.
The AS gravity parameters are set at $f_g=0.0096$, $f_y=0.00025$.
}
\label{fig:NO_plots}
\end{figure}

We show in \reffig{fig:NO_plots}(a) RGE trajectories producing 
the neutrino Yukawa couplings given in \refeq{spl_NO_H} and the mixing parameters given in \refeq{eq:phen_mix}. The flow of all other SM couplings is matched, after decoupling the gravitational contribution at $M_{\textrm{Pl}}$, to their low-scale value. Their effect (mostly the effect of $g_3$) is felt only in the IR, and it is thus not visible in the plots.  
In \reffig{fig:NO_plots}(b) we show a fit to \refeq{spl_NO} where, again, the mixing parameters are consistent with their experimental determination. While trajectories can be found for neutrino masses in NO for both the hierarchical and nearly degenerate cases of \refeq{spl_NO_H} and \refeq{spl_NO}, respectively, we are not able to find a single solution that matches the case with IO at the low scale. The parameter $X$ always appear to run to values that are much smaller than the lower bound  given in \refeq{eq:phen_mix}.

\section{Sterile-neutrino dark matter from freeze-in\label{sec:freezein}}

An intriguing feature of right-handed sterile neutrinos, besides providing the main ingredient for generating the 
Majorana mass of active neutrinos, is that they may 
comprise the observed dark matter abundance of the Universe if this is obtained via the freeze-in mechanism\cite{McDonald:2001vt,Choi:2005vq,Kusenko:2006rh,Petraki:2007gq,Hall:2009bx} (see, \textit{e.g.}, Refs.\cite{Shakya:2015xnx,Bernal:2017kxu} for reviews). As is well known, freeze-in provides a reliable process for dark matter production when the coupling between the visible sector and the dark matter is very small, so that the latter never reaches thermal equilibrium. The observed relic 
abundance is eventually generated by the decay or annihilation of the particles in the thermal bath
into dark-matter final states.     

Given that a feeble interaction between the particles in thermal 
equilibrium and the dark matter candidate is required, we use this section to extend the mechanism for a dynamical generation of small Yukawa couplings introduced in \refsec{sec:gen_mech} to a setting beyond the SM. We thus consider the simple model of Refs.\cite{Kusenko:2006rh,Petraki:2007gq} (see also Ref.\cite{Frigerio:2014ifa} for a slightly modified scenario), in which the 
sterile-neutrino dark matter that freezes in is the decay product of a gauge singlet scalar $S$ connected to one or more sterile 
neutrinos via a Yukawa coupling. Assuming for simplicity one generation of sterile neutrinos, $\nu_{R}$, 
one can write down the Lagrangian  
\be\label{eq:FIMP}
\mathcal{L}\supset -Y_{S}\, S\, \nu_{R} \nu_{R}- Y_L\, S \chi\, \xi +\textrm{H.c.} \,,
\ee
where we anticipate the presence of an additional Yukawa interaction of $S$ with two NP Weyl spinors $\xi, \chi$ of 
opposite dark (or SM) charge $-Q_D, Q_D$, 
which may be the (left-chiral) components of a Dirac fermion or else. 

If the scalar field $S$ decays to the sterile neutrino
while being in thermal equilibrium the dark matter relic abundance is given by\cite{Kusenko:2006rh,Petraki:2007gq}
\be
\Omega h^2 \approx 0.12 \left(\frac{Y_S}{10^{-8}} \right)^2 \left( \frac{M_N}{10^{-8}\, m_S} \right)\,,
\ee
where $m_S$ is the NP scalar mass and $M_N$ is the Majorana mass.
In the presence of a light (sub-MeV) sterile neutrino with feeble interactions to the visible sector, 
the correct relic abundance can be reached if the scalar 
is heavier than the dark matter by several orders of magnitude.    

One may apply the generic discussion presented in \refsec{sec:gen_mech} to the system composed of $Y_L$, $Y_S$, and $g_D$, where the latter is the dark U(1)$_D$ gauge coupling.\footnote{We assume that the parameters of the scalar potential can be safely excluded from the discussion. On the one hand, mass parameters are canonically relevant and do not provide predictions. On the other, quartic couplings 
do not enter the Yukawa RGEs at one loop. We also tacitly imply that AS quantum-gravity corrections to the RGEs of quartic couplings make them develop (pseudo-)Gaussian irrelevant fixed points,
following several works in the literature that have pointed to this conclusion\cite{Shaposhnikov:2009pv,Eichhorn:2017als,Pawlowski:2018ixd}. 
Provided that the SM quartic coupling is made approximately consistent with the Higgs mass value, the remaining portal 
couplings should be small enough to be factored out of the low energy phenomenology in a first approximation.}
We use \texttt{RGBeta}\cite{Thomsen:2021ncy} 
to generate the 1-loop RGEs 
of the model in \refeq{eq:FIMP}. 
In the trans-Planckian regime they read
\bea
\frac{d g_D}{dt}&=&\frac{g_D^3}{16 \pi^2}\frac{4}{3}Q_D^2 -f_g\, g_D\\
\frac{d Y_S}{dt}&=&\frac{Y_S}{16 \pi^2}\left(Y_L^2 +6\, Y_S^2+2\, y_\nu^2\right)-f_y\, Y_S\label{eq:NP_YS}\\
\frac{d Y_L}{dt}&=&\frac{Y_L}{16 \pi^2}\left(2\,Y_S^2 +2\, Y_L^2 -6\,Q_D^2\,g_D^2 \right)-f_y\, Y_L\\
\frac{d y_\nu}{dt}&=&\frac{y_\nu}{16 \pi^2}\left(\frac{5}{2}y_\nu^2+2\,Y_S^2 +3\, y_t^2 -\frac{3}{4}g_Y^2 \right)-f_y\, y_\nu\,,\label{eq:nuneu}
\eea
and the beta functions of $g_Y$ anf $y_t$ remain unchanged with respect to Eqs.~(\ref{eq:betagy}), (\ref{eq:betayt}).

A small $Y_S$ is dynamically generated if the trans-Planckian RGEs admit an IR-attractive fixed point, 
$g_D^{\ast}\neq 0$, $Y_L^{\ast}\neq 0$, $Y_S^{\ast}=0$. 
Referring to \refeq{eq:crit_fy}, with $X=L$, $Y=D$, $Z=S$, one can write down
\be
f_{S, L D}^{\textrm{crit}}=\frac{g_D^{\ast 2}}{16 \pi^2}\,\frac{\alpha_L' \alpha_D-\alpha_D' \alpha_L}{\alpha_L-\alpha_L'}\,,
\ee
where in this particular example $\alpha_L=2$, $\alpha_L'=1$, $\alpha_D=6 Q_D^2$, and $\alpha_D'=0$.

It follows from our discussion in \refsec{sec:gen_mech} that the condition for generating dynamically a small Yukawa coupling $Y_S$ is given by $f_y < f_{S, L D}^{\textrm{crit}}$. On the other hand, 
from our discussion in \refsec{sec:top-neu} we expect $f_y$ to assume a small ($\mathcal{O}[10^{-4}]$) positive value, so to achieve 
phenomenological consistency with the low-energy determinations of the top and bottom quark masses, and at the same time reduce dynamically the neutrino Yukawa coupling. As a rule of thumb, one might thus set 
 $f_{S, L D}^{\textrm{crit}}>0$, which produces the following general conditions:
\begin{itemize}
\item If $\alpha_L > \alpha_L'$ (like in the example given here), one should require 
\be
\alpha_D> \alpha_D' \frac{\alpha_L}{\alpha_L'}
\ee
\item If $\alpha_L < \alpha_L'$, the opposite relation applies, 
\be
\alpha_D< \alpha_D' \frac{\alpha_L}{\alpha_L'}\,.
\ee
\end{itemize}
 
\begin{figure}[t]
\centering
\includegraphics[width=0.44\textwidth]{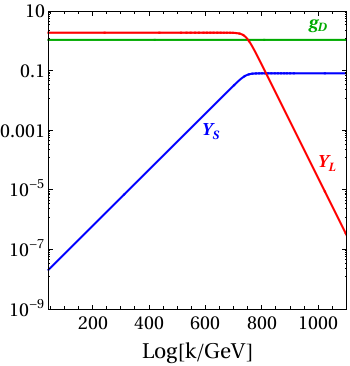}
\caption{The flow of the NP couplings $Y_L$ (solid red), $Y_S$ (solid blue) and dark gauge $g_D$ (solid green) in the trans-Planckian regime. Gauge charge and AS gravity parameters are set
at $Q_D=1$, $f_g=0.0096$, $f_y=0.00025$.}
\label{fig:Run_Fimp}
\end{figure}

The flow of the couplings $Y_L$ (solid red), $Y_S$ (solid blue) and dark gauge $g_D$ (solid green) in the trans-Planckian regime is 
presented in \reffig{fig:Run_Fimp}. We point out that the plot is not affected by the Yukawa couplings of the SM: the only one entering the running of $Y_S$ is in fact 
the neutrino Yukawa $y_\nu$, which is already of the order of $10^{-13}$ at the renormalization scale shown in  \reffig{fig:Run_Fimp}. 
Conversely, in the SM only \refeq{eq:nuneu} is modified by the $Y_S$ term which, however, does not interfere with the generation of a small  
neutrino Yukawa coupling because at the extreme UV scales presented in \reffig{fig:TM_Phase}(b) $Y_S$ remains at its interactive fixed point.

\section{Summary and conclusions\label{sec:summary}}

In this study we employed the framework of asymptotic safety above the Planck scale to investigate 
the possibility of generating in a dynamical way a Yukawa coupling of arbitrarily small size. This can emerge naturally
in a gauge-Yukawa system embedded in AS gravity if the Yukawa beta function develops a Gaussian fixed point along an irrelevant, IR-attractive direction of the RG flow. While a standalone fixed point of this kind would in general signal a ``triviality'' problem for the small coupling, we find that in the SM and other NP scenarios this is avoided by the presence of additional UV-attractive fixed points in the RGEs. The RG trajectories sprouting out of those UV fixed points eventually fall into the basin of attraction of the IR-attractive one, making the theory well defined from the Planck scale up to infinity.

We have applied this mechanism to two distinct but analogous scenarios. First, we have considered in some generality the case of neutrino mass generation in the SM. We have quantified the level of fine tuning (of the parameters associated with the strength of AS gravity) required to obtain neutrino masses of the Majorana or Dirac type at sub-Planckian scales, and we have identified the regions of the parameter space in which Dirac neutrino masses can be generated naturally thanks to the dynamical mechanism described above. Such solutions admit normal mass ordering and allow for the mixing parameters to be in agreement with current experimental determinations. 
In this context it is perhaps quite intriguing that a very 
recent analysis\cite{Jimenez:2022dkn} has found \textit{decisive} Bayesian evidence for the normal mass ordering based on the cosmological and oscillation experiment constraints on the neutrino masses.

We point out, however, that despite the potential for enhanced predictivity that in a fixed-point analysis is usually associated
with the presence of irrelevant RG flow directions, in the solutions that we found in agreement with low-scale observations 
the neutrino mixing parameters do not emerge as a prediction but rather stem from a completely relevant fixed point. 
This might be slightly disappointing but should not be entirely surprising. As gravity is flavor-blind by construction it is not in general expected to generate dynamically phenomena like the mixing of different generations of quarks and leptons, which are most likely due to a (still missing) UV theory of flavor.  

The second scenario we have considered is a model where sterile neutrinos comprise a 
light (sub-MeV) dark matter component of the Universe. As is well known, freeze-in provides in this case a viable mechanism for obtaining the correct relic density, if the dark matter particle interacts feebly with the visible sector, so to never reach thermal equilibrium.
The correct relic abundance emerges thanks to the decay of heavy particles in equilibrium with the thermal bath. We have shown that in the context of asymptotic safety the dynamical mechanism described above, based on the presence of an IR-attractive Gaussian fixed point, naturally yields Yukawa couplings of the expected size for the sterile-neutrino dark matter. In order to produce the additional UV-attractive fixed point(s) that renders the theory complete, one ought to introduce an abelian gauge interaction and a ``mirror'' Yukawa coupling with heavy particles, 
in similar fashion to the case of Dirac neutrinos in the SM, where the role of the mirror coupling is played by the top Yukawa term in the Lagrangian.      
Overall, the mechanism introduced in this study presents features that make it UV-complete, generic, and versatile enough to be successfully adopted in the context of other NP models with feeble Yukawa interactions.

\section*{Acknowledgments} 

The work of K.K. and S.P. is supported by the National Science Centre (Poland) under the research Grant No.~2017/26/E/ST2/00470. E.M.S. is supported in part by the National Science Centre (Poland) under the research Grant No.~2020/38/E/ST2/00126. The use of the CIS computer cluster at the National Centre for Nuclear Research in Warsaw is gratefully acknowledged.

\newpage
\appendix

\section{Renormalization group equations\label{app:appendix}}

In this appendix we present the trans-Planckian one-loop RGEs employed in \refsec{sec:SM} for the fixed-point analysis of the SM with right-handed neutrinos.
In both the quark and lepton sectors we keep the system in the mass basis along the entire flow. The rotation matrices become thus running parameters of the Lagrangian.

\subsection{Gauge, top, and bottom sector\label{app:gauge}}

Since the main focus of this work is the lepton sector, we simplify the RGEs of the quark sector down to their essential features. We work in the approximation of diagonal Yukawa matrices for the quarks, neglecting RG effects due to the running of the CKM matrix. We also limit our analysis to the third quark generation only. A thorough investigation of the trans-Planckian regime for the quark sector SM Yukawa and CKM matrix elements 
can be found in Ref.\cite{Alkofer:2020vtb}. Earlier investigations of the RG effects on the CKM matrix can be found, \textit{e.g.}, in Refs.\cite{Babu:1987im,Sasaki:1986jv,Barger:1992pk,Kielanowski:2008wm}.    

The RGEs for the gauge and top/bottom Yukawa sector of the SM read 
\be
\frac{d g_Y}{dt}=\frac{g_Y^3}{16\pi^2}\,\frac{41}{6}-f_g\, g_Y
\ee
\be
\frac{d g_2}{dt}=-\frac{g_2^3}{16\pi^2}\,\frac{19}{6}-f_g\, g_2
\ee
\be
\frac{d g_3}{dt}=-\frac{g_3^3}{16\pi^2}\,7-f_g\, g_3
\ee
\be
\frac{d y_t}{d t}=\frac{y_t}{16\pi^2}\left[\frac{9}{2} y_t^2+\frac{3}{2}y_b^2
-\left(\frac{17}{12} g_Y^2+\frac{9}{4} g_2^2 + 8 g_3^2  \right)
+y_e^2+y_\mu^2+y_\tau^2+y_{\nu 1}^2+y_{\nu 2}^2+y_{\nu 3}^2 \right]-f_y\, y_t
\ee
\be
\frac{d y_b}{dt}=\frac{y_b}{16\pi^2}\left[\frac{9}{2} y_b^2+\frac{3}{2} y_t^2
-\left(\frac{5}{12} g_Y^2+\frac{9}{4} g_2^2 + 8 g_3^2  \right)
+y_e^2+y_\mu^2+y_\tau^2+y_{\nu 1}^2+y_{\nu 2}^2+y_{\nu 3}^2 \right]-f_y\, y_b\,.
\ee

\subsection{Full lepton sector}

To investigate RG effects on the lepton sector we adapt to the PMNS matrix 
the parametrization introduced in Ref.\cite{Alkofer:2020vtb} for the CKM matrix. 

Given the PMNS matrix elements $U_{\alpha i}$, where $\alpha=e, \mu, \tau$ and $i=1, 2,3$ 
denote the charged leptons and the neutrinos, respectively,
we define $X=|U_{e1}|^2$,  $Y=|U_{e2}|^2$,  $Z=|U_{\mu 1}|^2$,  $W=|U_{\mu 2}|^2$
so that
\begin{eqnarray}\label{eq:thetas}
\theta_{12}&=& \arctan\sqrt{\frac{Y}{X}}\nonumber \\
\theta_{13}&=& \arccos\sqrt{X+Y}\nonumber \\
\theta_{23}&=& \arcsin\sqrt{\frac{1-W-Z}{X+Y}}\,,
\end{eqnarray}
and
\begin{equation}
\delta=\arccos\frac{(X+Y)^2Z-Y(X+Y+Z+W-1)-X(1-W-Z)(1-X-Y)}{2\sqrt{XY(1-X-Y)(1-Z-W)(X+Y+Z+W-1)}}\,,
\end{equation}
where the PMNS mixing angles are defined in Eqs.~(\ref{PMNS}), (\ref{3mixing}).

One can thus write the matrix of squared PMNS elements as
\begin{eqnarray}
U_2={|U_{\alpha i}|^2}=\left[
          \begin{array}{ccc}
         X & Y & 1-X-Y \\
 Z& W & 1-Z-W\\
  1-X-Z& 1-Y-W & X+Y+Z+W-1 \end{array} \right]
\;\;.
\label{PMNS2}
\end{eqnarray}
The allowed $3\,\sigma$ ranges of the parameters $X$, $Y$, $Z$, and $W$\cite{Esteban:2020cvm} are given in 
\refeq{eq:phen_mix}.

The full set of RGEs in the lepton sector reads, 
\begin{eqnarray}
\frac{d y_e}{d t}&=&\frac{y_e}{16\pi^2}\left\{ \frac{3}{2}y_e^2-\frac{3}{2}\left[X y_{\nu 1}^2 +Y y_{\nu 2}^2+(1-X-Y)y_{\nu 3}^2 \right]
+y_e^2+y_\mu^2+y_\tau^2+y_{\nu 1}^2+y_{\nu 2}^2+y_{\nu 3}^2 \right.\nonumber \\
&-&\left.\left(\frac{15}{4}g_Y^2+\frac{9}{4}g_2^2\right) 
+3\left( y_t^2+y_b^2\right)\right\}-f_y\, y_e
\end{eqnarray}

\begin{eqnarray}
\frac{d y_{\mu}}{dt}&=&\frac{y_\mu}{16\pi^2}\left\{ \frac{3}{2}y_\mu^2-\frac{3}{2}\left[Z y_{\nu 1}^2 +W y_{\nu 2}^2+(1-Z-W)y_{\nu 3}^2 \right]
+y_e^2+y_\mu^2+y_\tau^2+y_{\nu 1}^2+y_{\nu 2}^2+y_{\nu 3}^2 \right.\nonumber \\
&-&\left.\left(\frac{15}{4}g_Y^2+\frac{9}{4}g_2^2\right) 
+3\left(y_t^2+y_b^2\right)\right\}-f_y\,y_\mu
\end{eqnarray}

\begin{eqnarray}
\frac{d y_{\tau}}{dt}&=&\frac{y_\tau}{16\pi^2}\left\{ \frac{3}{2}y_\tau^2-\frac{3}{2}\left[(1-X-Z)y_{\nu 1}^2 + (1-Y-W) y_{\nu 2}^2+(X+Y+Z+W-1)y_{\nu 3}^2 \right]\right.\nonumber \\
&+&y_e^2+y_\mu^2+y_\tau^2+y_{\nu 1}^2+y_{\nu 2}^2+y_{\nu 3}^2
-\left.\left(\frac{15}{4}g_Y^2+\frac{9}{4}g_2^2\right) 
+3\left( y_t^2+y_b^2\right)\right\}-f_y\,y_\tau
\end{eqnarray}

\begin{eqnarray}\label{eq:y1rge}
\frac{d y_{\nu 1}}{dt}&=&\frac{y_{\nu 1}}{16\pi^2}\left\{ \frac{3}{2}y_{\nu 1}^2-\frac{3}{2}\left[X y_e^2 +Z y_\mu^2+(1-X-Z)y_\tau^2 \right]
+y_e^2+y_\mu^2+y_\tau^2+y_{\nu 1}^2+y_{\nu 2}^2+y_{\nu 3}^2 \right.\nonumber \\
&-&\left.\left(\frac{3}{4}g_Y^2+\frac{9}{4}g_2^2\right) 
+3\left( y_t^2+y_b^2\right)\right\}-f_y\, y_{\nu 1}
\end{eqnarray}

\begin{eqnarray}
\frac{d y_{\nu 2}}{dt}&=&\frac{y_{\nu 2}}{16\pi^2}\left\{ \frac{3}{2}y_{\nu 2}^2-\frac{3}{2}\left[Y y_e^2 +W y_\mu^2+(1-Y-W)y_\tau^2 \right]
+y_e^2+y_\mu^2+y_\tau^2+y_{\nu 1}^2+y_{\nu 2}^2+y_{\nu 3}^2 \right.\nonumber \\
&-&\left.\left(\frac{3}{4}g_Y^2+\frac{9}{4}g_2^2\right) 
+3\left( y_t^2+y_b^2\right)\right\}-f_y\, y_{\nu 2}
\end{eqnarray}

\begin{eqnarray}\label{eq:y3rge}
\frac{d y_{\nu 3}}{dt}&=&\frac{y_{\nu 3}}{16\pi^2}\left\{ \frac{3}{2}y_{\nu 3}^2-\frac{3}{2}\left[(1-X-Y) y_e^2 
+ (1-Z-W) y_\mu^2+(X+Y+Z+W-1)y_\tau^2 \right]\right.\nonumber \\
&+&y_e^2+y_\mu^2+y_\tau^2+y_{\nu 1}^2+y_{\nu 2}^2+y_{\nu 3}^2
-\left.\left(\frac{3}{4}g_Y^2+\frac{9}{4}g_2^2\right) 
+3\left(y_t^2+y_b^2\right)\right\}-f_y\,y_{\nu 3}
\end{eqnarray}

{\footnotesize
\begin{eqnarray}\label{eq:Xrge}
\frac{dX}{dt}&=&-\frac{3}{(4\pi)^2}\left[ 
\left( \frac{y_e^2+y_\mu^2}{y_e^2-y_\mu^2} \right) \left\{(y_{\nu 1}^2-y_{\nu 3}^2)XZ+\frac{(y_{\nu 3}^2-y_{\nu 2}^2)}{2}  [W(1-X)+X-(1-Y)(1-Z)] \right\} \right.\nonumber \\
&+&\left. \left( \frac{y_e^2+y_\tau^2}{y_e^2-y_\tau^2} \right) \left\{(y_{\nu 1}^2-y_{\nu 3}^2)X(1-X-Z)+\frac{(y_{\nu 3}^2-y_{\nu 2}^2)}{2}  [(1-Y)(1-Z)-X(1-2Y)-W(1-X)]  \right\}\right.\nonumber \\
&+&\left. \left( \frac{y_{\nu 1}^2+y_{\nu 2}^2}{y_{\nu 1}^2-y_{\nu 2}^2} \right) \left\{(y_e^2-y_\tau^2)XY+\frac{(y_\tau^2-y_\mu^2)}{2}  [W(1-X)+X-(1-Y)(1-Z)]  \right\}\right.\nonumber \\
&+&\left. \left( \frac{y_{\nu 1}^2+y_{\nu 3}^2}{y_{\nu 1}^2-y_{\nu 3}^2} \right) \left\{(y_e^2-y_\tau^2)X(1-X-Y)+\frac{(y_\tau^2-y_\mu^2)}{2}  [(1-Y)(1-Z)-X(1-2Z)-W(1-X)]  \right\}
\right]\nonumber \\
\end{eqnarray}

\begin{eqnarray}
\frac{dY}{dt}&=&-\frac{3}{(4\pi)^2}\left[ 
\left( \frac{y_e^2+y_\mu^2}{y_e^2-y_\mu^2} \right) \left\{\frac{(y_{\nu 3}^2-y_{\nu 1}^2)}{2}[W(1-X)+X-(1-Y)(1-Z)]+(y_{\nu 2}^2-y_{\nu 3}^2)YW \right\} \right.\nonumber \\
&+&\left. \left( \frac{y_e^2+y_\tau^2}{y_e^2-y_\tau^2} \right) \left\{\frac{(y_{\nu 3}^2-y_{\nu 1}^2)}{2}  [(1-Y)(1-Z)-W(1-X)-X(1-2Y)] +(y_{\nu 2}^2-y_{\nu 3}^2)Y(1-Y-W) \right\}\right.\nonumber \\
&+&\left. \left( \frac{y_{\nu 2}^2+y_{\nu 1}^2}{y_{\nu 2}^2-y_{\nu 1}^2} \right) \left\{(y_e^2-y_\tau^2)XY+\frac{(y_\tau^2-y_\mu^2)}{2}  [W(1-X)+X-(1-Y)(1-Z)]  \right\}\right.\nonumber \\
&+&\left. \left( \frac{y_{\nu 2}^2+y_{\nu 3}^2}{y_{\nu 2}^2-y_{\nu 3}^2} \right) \left\{(y_e^2-y_\tau^2)Y(1-X-Y)+\frac{(y_\mu^2-y_\tau^2)}{2}  [W(1-X-2Y)+X-(1-Z)(1-Y)]  \right\}
\right]\nonumber \\
\end{eqnarray}

\begin{eqnarray}
\frac{dZ}{dt}&=&-\frac{3}{(4\pi)^2}\left[ 
\left( \frac{y_\mu^2+y_e^2}{y_\mu^2-y_e^2} \right) \left\{(y_{\nu 1}^2-y_{\nu 3}^2)XZ +\frac{(y_{\nu 3}^2-y_{\nu 2}^2)}{2}[W(1-X)+X-(1-Y)(1-Z)]\right\} \right.\nonumber \\
&+&\left. \left( \frac{y_\mu^2+y_\tau^2}{y_\mu^2-y_\tau^2} \right) \left\{
(y_{\nu 1}^2-y_{\nu 3}^2)Z(1-X-Z)+\frac{(y_{\nu 2}^2-y_{\nu 3}^2)}{2}  [W(1-X-2Z)+X-(1-Y)(1-Z)] \right\}\right.\nonumber \\
&+&\left. \left( \frac{y_{\nu 1}^2+y_{\nu 2}^2}{y_{\nu 1}^2-y_{\nu 2}^2} \right) \left\{\frac{(y_e^2-y_\tau^2)}{2}  [(1-Y)(1-Z)-X-W(1-X)] + (y_\mu^2-y_\tau^2)ZW \right\}\right.\nonumber \\
&+&\left. \left( \frac{y_{\nu 1}^2+y_{\nu 3}^2}{y_{\nu 1}^2-y_{\nu 3}^2} \right) \left\{\frac{(y_\tau^2-y_e^2)}{2}  [(1-Z)(1-Y)-W(1-X)-X(1-2Z)]+(y_\mu^2-y_\tau^2)Z(1-Z-W)\right\}
\right]\nonumber \\
\end{eqnarray}

\begin{eqnarray}\label{eq:Wrge}
\frac{dW}{dt}&=&-\frac{3}{(4\pi)^2}\left[ 
\left( \frac{y_\mu^2+y_e^2}{y_\mu^2-y_e^2} \right) \left\{(y_{\nu 2}^2-y_{\nu 3}^2)WY +\frac{(y_{\nu 3}^2-y_{\nu 1}^2)}{2}[W(1-X)+X-(1-Y)(1-Z)]\right\} \right.\nonumber \\
&+&\left. \left( \frac{y_\mu^2+y_\tau^2}{y_\mu^2-y_\tau^2} \right) \left\{
(y_{\nu 2}^2-y_{\nu 3}^2) W(1-Y-W) + \frac{(y_{\nu 3}^2-y_{\nu 1}^2)}{2}  [(1-Y)(1-Z)-X-W(1-X-2Z)] \right\}\right.\nonumber \\
&+&\left. \left( \frac{y_{\nu 2}^2+y_{\nu 1}^2}{y_{\nu 2}^2-y_{\nu 1}^2} \right) \left\{(y_\mu^2-y_\tau^2)WZ + \frac{(y_\tau^2-y_e^2)}{2}  [(1-X)W+X-(1-Y)(1-Z)]  \right\}\right.\nonumber \\
&+&\left. \left( \frac{y_{\nu 2}^2+y_{\nu 3}^2}{y_{\nu 2}^2-y_{\nu 3}^2} \right) \left\{(y_\mu^2-y_\tau^2)W(1-Z-W) +\frac{(y_\tau^2-y_e^2)}{2}[(1-Y)(1-Z)-X-W(1-X-2Y)]\right\}
\right]\,.\nonumber \\
\end{eqnarray}}

\bigskip
\bibliographystyle{JHEP}
\bibliography{myref}

\end{document}